\documentclass{aa}

\usepackage{graphicx}
\usepackage{txfonts}
\usepackage{color}
\usepackage{xspace}
\usepackage{siunitx}
\usepackage{amsmath}
\usepackage{amssymb}
\usepackage{xcolor}
\usepackage{dashbox}
\usepackage{framed}
\usepackage{lipsum}
\usepackage{placeins}
\usepackage{stfloats}
\usepackage{upgreek}
\usepackage{lscape}
\usepackage{tikz}

\definecolor{purple}{HTML}{B76CE6}
\definecolor{lightblue}{HTML}{71A6D2}
\definecolor{brightgreen}{HTML}{65E67D}
\definecolor{green}{HTML}{A6D854}
\definecolor{yellow}{HTML}{E6C565}
\definecolor{red}{HTML}{DC143C} 
\definecolor{edgegray}{HTML}{696969}

\usepackage{hyperref}

\hypersetup{
        colorlinks=true, 
        breaklinks=true,
        linkcolor=blue, 
        citecolor=blue, 
        filecolor=blue, 
        urlcolor=blue,
        unicode=false, 
        pdftoolbar=true, 
        pdfmenubar=true, 
        pdffitwindow=false, 
        pdfstartview={Fit}, 
        pdftitle={Mapping water ice with infrared broadband photometry}, 
        pdfauthor={Stefan Meingast}, 
        pdfsubject={Mapping water ice with infrared broadband photometry},
        pdfcreator={Stefan Meingast}, 
        pdfkeywords={},
        pdfnewwindow=true, 
        pdfdisplaydoctitle=true 
}

\makeatletter
\renewcommand*\aa@pageof{, page \thepage{} of \pageref*{LastPage}}
\makeatother

\newcommand{\ice}{\textsc{Ice}\xspace}
\newcommand{\taumax}{\ensuremath{\tau_{3.0}^{\max}}\xspace}
\newcommand{\lambdaWISEminusIRAC}{\ensuremath{\Lambda (W_1 - I_1)}\xspace}

\DeclareSIUnit{\jansky}{Jy}

\begin{document}

\title{Mapping water ice with infrared broadband photometry}

\author{Stefan Meingast\inst{1}}

\institute{University of Vienna, Department of Astrophysics, T\"urkenschanzstrasse 17, 1180 Wien, Austria 
\\ \email{stefan.meingast@univie.ac.at}
}

\date{Received May 16, 2025 / Accepted July 24, 2025}

\abstract{Interstellar ices play a fundamental role in the physical and chemical evolution of molecular clouds and star-forming regions, yet their large-scale distribution and abundance remain challenging to map. In this work, I present the ice color excess method (\ice), which parametrizes the peak optical depth (\taumax) of the prominent \SI{3}{\micro\meter} absorption feature, which is predominantly caused by the presence of solid H\(_2\)O. The method builds on well-established near-infrared color excess techniques and uses widely available infrared broadband photometry. Through detailed evaluation of passband combinations and a comprehensive error analysis, I construct the \ice color excess metric \lambdaWISEminusIRAC. This parameter emerges as the optimal choice that minimizes systematic errors while leveraging high-quality, widely available photometry from \textit{Spitzer} and \textit{WISE} data archives. To calibrate the method, I compile from the literature a sample of stars located in the background of nearby molecular clouds, for which spectroscopically measured optical depths are available. The empirical calibration yields a remarkably tight correlation between \taumax and \lambdaWISEminusIRAC. This photometric technique opens a new avenue for tracing the icy component of the interstellar medium on Galactic scales, providing a powerful complement to spectroscopic surveys and enables new insights into the environmental dependence of the formation and evolution of icy dust grains.}


\maketitle

\section{Introduction}
\label{sec:introduction}

Interstellar ices are a key component of the coldest and densest phases of the interstellar medium, profoundly influencing both the physical and chemical evolution of molecular clouds and the processes of star and planet formation \citep{Boogert2015}. One of the most prominent tracers of these ices is the broad \SI{3}{\micro\meter} absorption feature, which arises primarily from the O--H stretching mode in solid H\(_2\)O, but whose profile is shaped by additional constituents such as CH\(_3\)OH and NH\(_3\)\(\cdot\)H\(_2\)O and by the physical structure of icy grain mantles \citep[e.g.,][]{Tielens1984, Baratta1990, Boogert2011, Noble2013, McClure2023}. This feature not only signals the presence of water ice but also encodes information about the composition, environment, and evolutionary history of dust grains in molecular clouds.

Over the past decades, tremendous observational and theoretical efforts have been devoted to unraveling the complexity of interstellar ices and their composition \citep[see e.g.,][and references therein]{Hudgins1993, Boogert2015, Dartois2024}. Laboratory experiments and astrochemical modeling have revealed that interstellar ices are not simple, pure substances but rather complex mixtures whose spectral features are sensitive to temperature, composition, and grain morphology \citep{Hudgins1993, Baratta1990, Smith1993}. Observationally, the \SI{3}{\micro\meter} feature has been a cornerstone for tracing ice in the Galaxy, but its study has been hindered by significant technical challenges. Ground-based observations in this wavelength regime are hampered by strong telluric absorption, requiring exceptionally careful calibration and data reduction \citep[e.g.,][]{Danielson1965,Knacke1969,Gillet1973, Murakawa2000, Chiar2011}. Even when high-quality spectra can be obtained, deriving accurate optical depths for the \SI{3}{\micro\meter} feature is complicated by uncertainties in the underlying dust extinction law, the spectral type of the background star, and the presence of overlapping features from other species \citep[e.g.,][]{Boogert2011, Boogert2013, Madden2022}.

The advent of space-based observatories has dramatically advanced the field. The \textit{Spitzer} Space Telescope \citep{Werner2004,Fazio2004}, with its Infrared Spectrograph \citep[IRS;][]{Houck2004}, has enabled systematic studies of ice absorption features in a wide range of environments, providing a wealth of high-quality spectra that have shaped our current understanding of ice chemistry and distribution \citep[e.g.,][]{Boogert2011, Boogert2013, Madden2022}. More recently, the James Webb Space Telescope \citep[\textit{JWST};][]{Gardner2006,Greene2017,Jakobsen2022} has begun to deliver unprecedented sensitivity and spectral resolution in the mid-infrared, revealing new details in the structure and composition of interstellar ices and enabling studies of even fainter and more embedded sources \citep{McClure2023, Dartois2024, Rocha2025, Smith2025}.

Looking ahead, the Spectro-Photometer for the History of the Universe, Epoch of Reionization, and Ices Explorer mission \citep[\textit{SPHEREx};][]{Crill2020} will provide the first all-sky near-infrared spectral survey, systematically mapping ice absorption features such as H\(_2\)O, CO, and CO\(_2\). This promises a statistical revolution in our understanding of the Galactic ice reservoir and its environmental dependence.

Despite these advances and in anticipation of results from \textit{SPHEREx}, most current approaches to ice mapping rely on spectroscopic data for individual sources, which can be observationally expensive and limited in spatial coverage or depth. In this manuscript, I present a new approach: a parameterization of the peak optical depth, of the \SI{3}{\micro\meter} ice absorption feature (hereinafter referred to as \(\taumax\)) based on broadband infrared photometry. By leveraging photometric data archives from surveys such as the Wide-Field Infrared Survey Explorer \citep[\textit{WISE};][]{Wright2010} and \textit{Spitzer}, and by calibrating the method with a carefully constructed literature sample, I demonstrate that it is possible to robustly and efficiently trace interstellar ices on Galactic scales.

\section{The ice color excess}
\label{sec:ice}

The presence of water ice and other frozen constituents on dust grains introduces absorption features into the infrared spectra of background stars. Quantifying the additional extinction caused by these icy grains -- beyond common dust extinction -- is essential for tracing the distribution and abundance of interstellar ices. In this section, I introduce the concept of the ice color excess method \ice, review its physical basis and observational motivation, and present two definitions used to isolate and measure the contribution of ices to the observed extinction.

\subsection{Physical background and motivation}
\label{sec:motivation}

The O--H stretch in H\(_2\)O and \(\text{CH}_3\text{OH}\), as well as the N--H stretch in the \(\text{NH}_3\!\cdot\!\mathrm{H}_2\mathrm{O}\) hydrate, produce a broad absorption band approximately \SI{1}{\micro\meter} wide, peaking at around \SI{3}{\micro\meter}. Consequently, and similar to dust extinction, this additional absorption component significantly affects photometric measurements at these wavelengths.

\begin{figure}[t]
        \centering
        \resizebox{1.0\hsize}{!}{\includegraphics[]{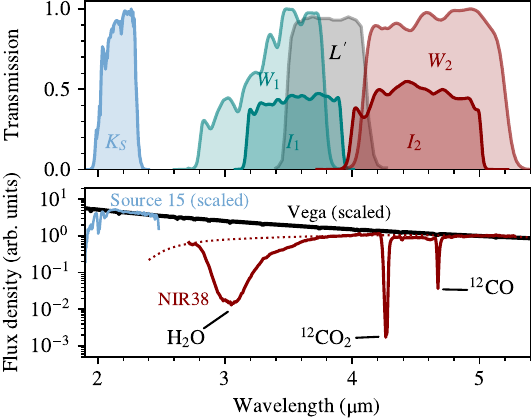}}
        \caption[]{Infrared passbands and spectral energy distributions (SEDs). The upper panel illustrates transmission curves for the infrared passbands \(K_S\), \(W_1\), \(I_1\), \(L'\), \(I_2\), and \(W_2\). The lower panel presents the SEDs of three distinct sources: the light blue line represents an X-SHOOTER spectrum of source 15 (refer to Table~\ref{tab:master}), the solid red line corresponds to the source NIR38, and the black line depicts Vega. All spectra have been rescaled to facilitate a better comparison.}
    \label{fig:spectra_passbands}
\end{figure}

Figure~\ref{fig:spectra_passbands} presents a selection of infrared passbands (see Sect.~\ref{sec:data} for further details) and spectral energy distributions (SEDs) for three distinct stellar sources. The upper panel displays the transmission curves of widely used passbands\footnote{In this manuscript, I refer to the first (i.e., bluest) passbands of the \textit{WISE} and \textit{Spitzer} missions as \(W_1\) and \(I_1\), corresponding to the WISE1 and IRAC1 filters, respectively. Details on passbands and transmission curves are taken from the Spanish Virtual Observatory (SVO) Filter Profile Service \citep{svo2012, svo2020}.}, ordered by increasing effective wavelength: \(K_S\), \(W_1\), \(I_1\), \(L'\), \(I_2\), and \(W_2\). The \(K_S\) passband \citep{Persson1998} is widely used in ground-based surveys such as the Two Micron All Sky Survey \citep[2MASS;][]{Skrutskie2006}, while the \(L'\) filter \citep[e.g.,][]{Simons2002} is also employed in ground-based observations, though it is less commonly available. The \(W_1\) and \(W_2\) passbands were used by the Wide-field Infrared Survey Explorer \citep[WISE;][]{Wright2010}, and \(I_1\) and \(I_2\) by the \textit{Spitzer} Space Telescope.

Specifically, three passbands cover (parts of) the broad absorption profile: $W_1$ covers wavelengths from approximately \SI{2.75}{\micro\meter} to \SI{3.87}{\micro\meter}, with an effective wavelength of \SI{3.35}{\micro\meter}. Similarly, the $I_1$ passband covers wavelengths from \SI{3.13}{\micro\meter} to \SI{3.96}{\micro\meter}, with an effective wavelength of \SI{3.51}{\micro\meter}. The \(L'\) passband, as for example used in the (now decommissioned) VLT NACO instrument, ranges from approximately 3.4 to \SI{4.3}{\micro\meter} and has an effective wavelength of \SI{3.8}{\micro\meter}.

The lower panel of Fig.~\ref{fig:spectra_passbands} shows SEDs -- specifically the flux density as a function of wavelength -- for three stellar sources, each rescaled for visualization. The spectrum of NIR38, a star in the background of the Chamaeleon I molecular cloud observed with \textit{JWST} \citep{McClure2023}, is shown in red, rescaled to a median flux density of \SI{1}{\milli\jansky} between 4.8 and \SI{5.2}{\micro\meter}. The black line represents the spectrum of Vega from the STScI CALSPEC database \citep{Bohlin2014}, rescaled in the same way as NIR38. The light blue line shows the spectrum of source 15 (see Table~\ref{tab:master}), obtained with the X-SHOOTER instrument \citep{Vernet2011} under ESO program ID 090.C-0050(A), and rescaled to match Vega between 2.0 and \SI{2.4}{\micro\meter}. Among all sources in the final selection (see Sect.~\ref{sec:optical_depth_data}), this was the only one with a NIR spectrum available in the ESO Science Portal\footnote{\href{https://archive.eso.org/scienceportal}{https://archive.eso.org/scienceportal}}.

Given the broad \SI{3}{\micro\meter} absorption feature, it is evident that photometric measurements of sources in the background of ice-rich dust clouds are strongly affected by its presence. In particular, the \(W_1\) and \(I_1\) passbands coincide with the deepest parts of the absorption feature. However, only the \(W_1\) filter fully covers the peak of the optical depth spectrum located at about \SI{3.0}{\micro\meter}. The \(W_2\) and \(I_2\) passbands are influenced by absorption from icy grain constituents such as \(^{12}\)CO\(_2\) and \(^{12}\)CO \citep[e.g.,][]{McClure2023}. In contrast, the \(K_S\) passband remains unaffected by ice-related extinction, making it a robust tracer for extinction caused by dust grains at these wavelengths, independent of the presence of ices. 

Given the characteristics of these passbands, Fig.~\ref{fig:spectra_passbands} demonstrates that photometric measurements with filters covering the broad \SI{3}{\micro\meter} ice absorption profile are significantly affected by an ice extinction component in addition to dust attenuation. Consequently, the ice extinction component -- superimposed on dust extinction -- can theoretically be measured in a manner analogous to established dust extinction techniques.

\subsection{Color excess approach}
\label{sec:definition}

Arguably, the most popular method to map dust extinction in the Galaxy is the Near-Infrared Color Excess method (\textsc{Nice}; \citealt{Lada1994}). This method, along with its successors \textsc{Nicer} \citep{Lombardi2001}, \textsc{Pnicer} \citep{Meingast2017}, and \textsc{Xnicer} \citep{Lombardi2018}, relates the measured photometric color excess of stars to the column density of (molecular) hydrogen. In general, the color excess is defined as 
\begin{equation}
E(m_1 - m_2) = (m_1 - m_2) - (m_1 - m_2)_0 = A_{m_1} - A_{m_2},    
\end{equation}
where \(m_1\) and \(m_2\) refer to flux measurements in magnitudes in two passbands, \(\left( m_1 - m_2 \right)_0\) denotes the intrinsic color of a source and \(A_{m_1}\) and \(A_{m_2}\) refer to the amount of extinction in these passbands also in magnitudes. Since their inception and through subsequent advancements, the methods to determine dust extinction have proven to be robust estimators of density information in the interstellar medium \citep[e.g.,][]{Goodman2009,Lombardi2014,Meingast2018,Zhang2022}.

In the context of extinction by icy grains, this concept can be extended further. Given an observed color
\begin{equation}
    \left( m_1 - m_2 \right) = \left( m_1 - m_2 \right)_0 + E(m_1 - m_2),
\end{equation}
the color excess can be decomposed into distinct dust and ice components:
\begin{equation}
    \left( m_1 - m_2 \right) = \left( m_1 - m_2 \right)_0 + E(m_1 - m_2)_{\rm dust} + E(m_1 - m_2)_{\rm ice}.
\end{equation}
This equation defines the principal observable of the \ice method:
\begin{align}
    \Lambda(m_1 - m_2) &\equiv E(m_1 - m_2)_{\rm ice} \\
    &= \left( m_1 - m_2 \right) - \left( m_1 - m_2 \right)_0 - E(m_1 - m_2)_{\rm dust}.
\end{align}
This formulation shows that the color excess due to ices can be measured if both the intrinsic color of the star and the extinction caused by dust along the line of sight can be independently determined.

However, both dust and ice extinction are encoded in the single observed color value. Therefore, it becomes necessary to estimate the dust extinction using passbands that are not affected by absorption arising from the presence of ices. Given the well-established dust extinction techniques at near-infrared wavelengths and the absence of significant ice absorption features in the \(K_S\) passband (see Fig.~\ref{fig:spectra_passbands}), \(E(m_1 - m_2)_{\rm dust}\) at wavelengths affected by ice can be estimated by extrapolating the extinction law. A practical implementation for measuring \(\Lambda\) is given by
\begin{equation}
\label{equ:ice}
    \Lambda(m_1 - m_2) = \left( m_1 - m_2 \right) - \left( m_1 - m_2 \right)_0 - A_{K_S} \cdot \Psi(m_1, m_2),
\end{equation}
where \(\Psi\) can be determined from the shape of the extinction law:
\begin{equation}
\label{equ:psi}
    \Psi(m_1, m_2) = \frac{E(m_1 - m_2)_{\rm dust}}{A_{K_S}}.
\end{equation}
The term \(\left( m_1 - m_2 \right)\) can be obtained from photometric measurements, the intrinsic color \(\left( m_1 - m_2 \right)_0\) must be measured or modeled, \(A_{K_S}\) can be computed from passbands unaffected by ice extinction (a practical example is near-infrared \(JHK_S\) photometry), and \(\Psi(m_1, m_2)\) can be derived from theoretical or empirically deduced extinction curves.

\subsection{Reddening-free index approach}
\label{sec:alt_definition}

The procedure described in the preceding section for measuring the ice color excess requires knowledge of the absolute extinction caused by interstellar dust. Alternatively, it is theoretically possible to define a reddening-free color metric that does not rely on a direct measurement of dust extinction. This concept originates from the work of \citet{Johnson1953}, who introduced the quantity
\begin{equation}
    Q = (U-B) - \frac{E(U-B)}{E(B-V)} \cdot (B-V),
\end{equation}
where \(Q\) is constructed to be independent of dust reddening. However, this definition does not account for intrinsic stellar colors.

Expanding on this idea, one can define a dust-reddening-free index, \(Q'\), that also removes the dependence on intrinsic colors:
\begin{equation}
\label{equ:ice_alt}
    Q' \equiv E(m_1 - m_2) - \frac{E(m_1 - m_2)_{\rm dust}}{E(m_2 - m_3)_{\rm dust}} \cdot E(m_2 - m_3),
\end{equation}
where \(m_1\), \(m_2\), and \(m_3\) are magnitudes in three different passbands.

For notational simplicity, let \(E(m_1 - m_2) = E_{12}\) and recall that \(E_{12} = A_{m_1} - A_{m_2}\), with \(A\) representing extinction in magnitudes. Equation~\eqref{equ:ice_alt} can then be rearranged as:
\begin{align}
    Q' &= E_{12, \rm dust} + E_{12, \rm ice} - \frac{E_{12, {\rm dust}}}{E_{23, \rm dust}} \cdot \left( E_{23, \rm dust} + E_{23, \rm ice} \right) \\
    &= E_{12, \rm dust} - \frac{E_{12, \rm dust}}{E_{23, \rm dust}} \cdot E_{23, \rm dust} + E_{12, \rm ice} - \frac{E_{12, \rm dust}}{E_{23, \rm dust}} \cdot E_{23, \rm ice} \\
    &= A_{m_1, \rm ice} - A_{m_2, \rm ice} - \frac{E_{12, \rm dust}}{E_{23, \rm dust}} \cdot \left( A_{m_2, \rm ice} - A_{m_3, \rm ice} \right). \label{equ:ice_alt1}
\end{align}

In the optimal case -- where the passbands \(m_1\) and \(m_3\) do not overlap with any extinction features originating from ices -- Eq.~\eqref{equ:ice_alt1} reduces to
\begin{equation}
    Q' = - A_{m_2, \rm ice} \cdot \left[1 + \frac{E(m_1 - m_2)_{\rm dust}}{E(m_2 - m_3)_{\rm dust}} \right].
\end{equation}
For a case where passband \(m_3\) also covers the \SI{3}{\micro\meter} feature Eq.~\eqref{equ:ice_alt1} becomes
\begin{equation}
\label{equ:ice_alt2}
    Q' = A_{m_3, \rm ice} \cdot \frac{E(m_1 - m_2)_{\rm dust}}{E(m_2 - m_3)_{\rm dust}} - A_{m_2, \rm ice} \cdot \left[1 + \frac{E(m_1 - m_2)_{\rm dust}}{E(m_2 - m_3)_{\rm dust}} \right]
\end{equation}
dampening the signal strength. 

Similar to the color excess approach, the practical implementation can follow a relatively straightforward approach: \(Q'\) can be determined using Eq.~\eqref{equ:ice_alt}, given that intrinsic colors and the dust extinction law are known.

\section{Sources of uncertainty}
\label{sec:errors}

Having outlined the methodology for the \ice method in Sects.~\ref{sec:definition} and \ref{sec:alt_definition} with two separate approaches, it is essential to consider the various sources of error that affect measurements in practice. Random errors are primarily associated with the precision of stellar flux measurements. In this regard, advances in instrumentation and the advent of large-scale surveys and homogeneous data processing techniques have reduced these uncertainties to the millimagnitude level for millions of sources across the sky. As a result, the dominant sources of uncertainty in measuring the ice color excess are systematic in nature.

The following sections address the most significant systematic error sources: uncertainties in the dust extinction law, intrinsic stellar colors, source variability, and variations in the shape of the \SI{3}{\micro\meter} ice absorption profile.

\subsection{Dust extinction law}
\label{sec:errors_extinction_law}

\begin{figure}[t]
        \centering
        \resizebox{1.0\hsize}{!}{\includegraphics[]{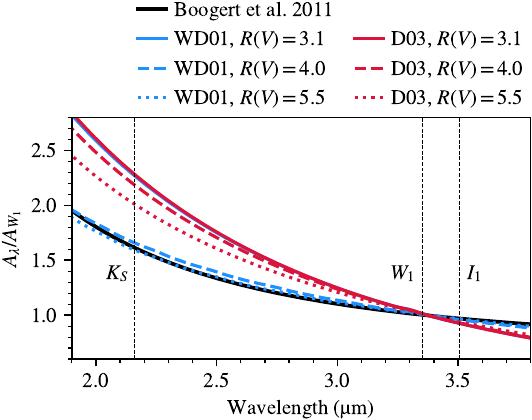}}
        \caption[]{Extinction curves normalized to their value at the effective wavelength of the \(W_1\) passband. The different lines correspond to various dust models and \(R(V)\) values, as indicated in the legend. WD01 refers to the models of \citet{Weingartner2001}, and D03 to those of \citet{Draine2003}. The curves illustrate how both the choice of dust model and \(R(V)\) parameter affect the wavelength dependence of extinction.}
    \label{fig:extinction_laws}
\end{figure}

To accurately determine the color excess attributable to interstellar ices, it is essential to disentangle the extinction caused by ices from that produced by dust. As discussed in Sect.~\ref{sec:ice}, this process requires adopting a dust extinction law that characterizes the relative contributions to extinction across different passbands. However, the precise form of the extinction law is often uncertain and can vary not only between different lines of sight, but also along a single line of sight. To assess the systematic impact of this uncertainty on ice color excess measurements, I compare several commonly used extinction laws, examining how their variations affect the results. The dust extinction laws considered here are those of \citet[WD01]{Weingartner2001}, \citet[D03]{Draine2003}, and \citet{Boogert2011}. For both WD01 and D03, I include curves corresponding to \(R(V) = A_V / E(B-V)\) values of 3.1, 4.0, and 5.5.

Figure~\ref{fig:extinction_laws} displays these extinction laws over the wavelength range from 1.9 to \SI{3.8}{\micro\meter}, normalized to their values at the effective wavelength of the \(W_1\) passband. The plot demonstrates that the extinction law flattens toward the mid-infrared. This is because, for dust grains much smaller than the wavelength of light, extinction is dominated by absorption. As grains grow to \si{\micro\meter} sizes, however, scattering becomes increasingly important in the infrared, resulting in a noticeable flattening of the extinction curve at wavelengths greater than about \SI{3}{\micro\meter} \citep[e.g.,][]{Dartois2024}. Consequently, the difference between extinction laws in the \(W_1\) and \(I_1\) passbands is much smaller than for pairs involving, for example, the \(K_S\) band at about \SI{2.2}{\micro\meter}. For instance -- given the set of extinction laws listed above -- \(A_{K_S} / A_{W_1}\) ranges from 1.6 to 2.3 -- a difference of about \SI{40}{\percent} between the various curves—whereas \(A_{W_1} / A_{I_1}\) varies by only about \SI{6}{\percent}, from 1.04 to 1.1.

More specifically, among the extinction laws considered, the factor \(\Psi(K_S, W_1)\) -- as defined in Eq.~\eqref{equ:psi} -- ranges from 0.37 to 0.56, representing a variation of approximately \SI{50}{\percent}. In contrast, \(\Psi(W_1, I_1)\) exhibits a smaller variation of about \SI{30}{\percent}, ranging from 0.03 to 0.039. These substantial uncertainties highlight the importance of the absolute values of \(\Psi(m_1, m_2)\) in the error budget. For a given value of \(A_{K_S}\), the final term in Eq.~\eqref{equ:ice} will have a larger impact on \(\Lambda\) when passbands with greater wavelength separations are used. Similarly, for the dust-reddening-free index \(Q'\) (see Sect.~\ref{sec:alt_definition}), the ratio \(E(m_1 - m_2) / E(m_2 - m_3)\) can significantly influence the total uncertainty. For the specific case of \(m_1 = K_S\), \(m_2 = W_1\), and \(m_3 = I_1\), the different extinction laws yield absolute values between 13.2 and 16.8 (a difference of roughly \SI{30}{\percent}), thus contributing notably to the uncertainty in Eq.~\eqref{equ:ice_alt}. For \(m_1 = K_S\), \(m_2 = W_1\), and \(m_3 = L'\), the corresponding values range from 5.7 to 7.6 (a difference of about \SI{35}{\percent}).

This analysis demonstrates that the choice of passband combination plays a crucial role in mitigating these uncertainties: selecting filters with closely spaced wavelengths, such as \(W_1\) and \(I_1\), substantially reduces the impact of extinction law variations on the error budget. Conversely, using passbands that are widely separated in wavelength can lead to much larger systematic uncertainties. Careful selection of filter pairs is therefore essential for robust measurements, particularly when the extinction law is not well constrained. The choice of passbands will be discussed in more detail in Sect.~\ref{sec:choice_passbands}.

\subsection{Intrinsic colors}
\label{sec:errors_intrinsic_colors}

Another critical aspect of measuring color excess is the accurate determination of intrinsic stellar colors. Fortunately, the spectral energy distribution of stars is relatively flat at infrared wavelengths, as this part of the spectrum lies in the Rayleigh-Jeans regime. As a result, the intrinsic colors of stars in infrared passbands exhibit a much narrower distribution compared to optical wavelengths. This characteristic is one of the main reasons why dust extinction measurements are particularly effective in the infrared \citep[e.g.,][]{Majewski2011}. 

Nevertheless, determining intrinsic colors remains a challenge, especially in the context of deep, modern surveys. In particular, contamination from extragalactic sources can complicate the color distribution and must be carefully addressed when developing methods to compute color excess \citep[e.g.,][]{Meingast2017}. For extinction mapping based on large samples, a rough estimate of the intrinsic color distribution is often sufficient, as the results are averaged over many sources. However, in this work I aim to parametrize \taumax using individual stellar colors. Consequently, it is essential to determine the intrinsic color for each star in the sample with high accuracy, and to quantify how uncertainties in intrinsic colors contribute to the overall error budget.

\begin{figure}[t]
        \centering
        \resizebox{1.0\hsize}{!}{\includegraphics[]{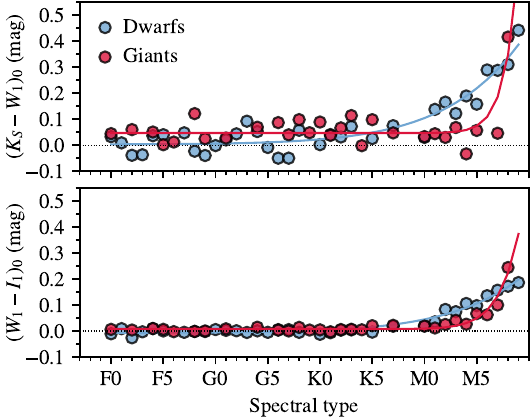}}
        \caption[]{Intrinsic colors for dwarfs (blue symbols) and giants (red symbols) as a function of spectral type, derived from synthetic photometry of IRTF spectra. Solid lines indicate power-law fits to the intrinsic colors for dwarfs (blue) and giants (red), respectively. The upper panel shows intrinsic \(K_S - W_1\) color, while the lower panel displays \(W_1 - I_1\). For both color indices, significant deviations from \SI{0}{mag} are observed only for late spectral types (later than approximately M0).}
    \label{fig:intrinsic_colors}
\end{figure}

To assess the impact of intrinsic color uncertainties for known spectral types, I computed intrinsic colors using observed stellar spectra from the IRTF Spectral Library\footnote{\href{https://irtfweb.ifa.hawaii.edu/~spex/IRTF_Spectral_Library/}{https://irtfweb.ifa.hawaii.edu/\string~spex/IRTF\_Spectral\_Library/}}. Synthetic photometry was performed with the PYPHOT package \citep{pyphot} on IRTF spectra spanning spectral types F0 to M9, for both dwarfs and giants. The analysis focused on the color indices most relevant to this study: \((K_S - W_1)_0\) and \((W_1 - I_1)_0\). Figure~\ref{fig:intrinsic_colors} visualizes the results: the top panel shows \((K_S - W_1)_0\) as a function of spectral type, while the bottom panel displays \((W_1 - I_1)_0\). Results for dwarfs and giants are indicated by blue and red filled circles, respectively. To mitigate noise and systematic errors in individual spectra, I applied least-squares power-law fits of the form \((m_1 - m_2)_0 = a \cdot x^b + c\), where \(x\) denotes the spectral type sequence from F0 to M9. The figure demonstrates that intrinsic colors remain close to zero for most spectral types, with significant deviations appearing only for late-type stars. This is particularly relevant, as M-type stars are the most numerous in the Galaxy, and late-type giants are often targeted in studies of background stars behind molecular clouds due to their high intrinsic luminosity, which enables detection through large amounts of extinction. I note that the apparent deviation for spectral type M8 for the giant sequence in \((K_S - W_1)_0\) does not affect the analysis, as this color is not used in subsequent steps. The smoothed intrinsic color values derived from the fits are listed in Table~\ref{tab:intrinsic_colors}, including values for the \((W_1 - L')_0\) color for future reference. I use the values obtained from the power-law fits for each spectral type to determine intrinsic colors for individual stars during the calibration of the method in Sect.~\ref{sec:calibration}.

To validate the synthetic photometry on IRTF spectra with observational data, I also extracted spectral type information from Simbad\footnote{\href{https://simbad.cds.unistra.fr/simbad/}{https://simbad.cds.unistra.fr/simbad/}}. To minimize the effects of extinction, I queried for all sources with \textit{Gaia} \citep[specifically data release 3;][]{Gaia,GaiaDR3} parallaxes greater than \SI{1}{mas}, while young stellar objects, variable stars, and binaries (as identified in Simbad) were excluded. Moreover, I required a cross-match with the unWISE and SESNA datasets (see Sect.~\ref{sec:infrared_photometry}). Due to these selection criteria, the sample of stars with luminosity class III  was too small to yield statistically significant results, and the comparison is therefore restricted to dwarfs (\(n=733\) for \(K_S - W_1\) and \(n=462\) for \(W_1 - I_1\)). In the same way as described above, I determined power-law fits to colors for these stars as a function of listed spectral type and compared the result to the synthetic photometry. My findings from this experiment are in excellent agreement with those derived from the IRTF spectra: the mean deviation is only \SI{0.02}{mag} in \(K_S - W_1\) and at the \si{mmag} level for \(W_1 - I_1\). Only for very late spectral types (later than approximately M5) does the comparison in \(W_1 - I_1\) reveal a systematic deviation of about \SI{0.05}{mag}. The comparison for dwarfs between the synthetic photometry from IRTF spectra and the Simbad database is visualized in Fig.~\ref{fig:intrinsic_colors_sed_vs_simbad}.

\subsection{Source variability}
\label{sec:errors_variablity}

Another source of systematic uncertainty is the intrinsic variability of stars. In this study, the photometry used to compute color excesses is derived from observations taken at different epochs, sometimes separated by more than a decade. For example, the 2MASS survey collected data between 1997 and 2001, the c2d survey lists \(I_1\) recording dates from 2003 to 2005, and \textit{WISE} observed from 2009 to 2024 (including NEOWISE; \citealt{Mainzer2011}). Additionally, some targeted observations of sources to measure the depth of the \SI{3}{\micro\meter} ice feature even predate the 2MASS survey. Since nearly all stars discussed in this paper are giants -- stellar types that are often long-period variables with timescales of several years -- long-term variability and its potential impact on the results must be carefully considered.

For ensembles of stars, variability is generally less problematic, as the sample mean flux averages out individual stellar fluctuations. However, for individual sources, it is essential to account for variability explicitly. To assess the impact of source variability, I utilize the unTimely source catalog \citep{Meisner2023}, which was constructed by measuring fluxes on a series of unWISE coadds in the \(W_1\) and \(W_2\) passbands, each corresponding to a biannual \textit{WISE} sky pass. With data spanning from 2010 to 2020, unTimely provides up to 16 measurements per detected source. Given only two data points per year, short-term variations are not well sampled, but the decade-long baseline is well suited to reveal long-term or large-amplitude variability. For the calibration of the method, I use the obtained information on variability from unWISE to construct a quality filter and to add an additional systematic error term in regard to photometric parameters.

\subsection{Shape of the absorption profile}
\label{sec:errors_profile}

\begin{figure}[t]
        \centering
        \resizebox{1.0\hsize}{!}{\includegraphics[]{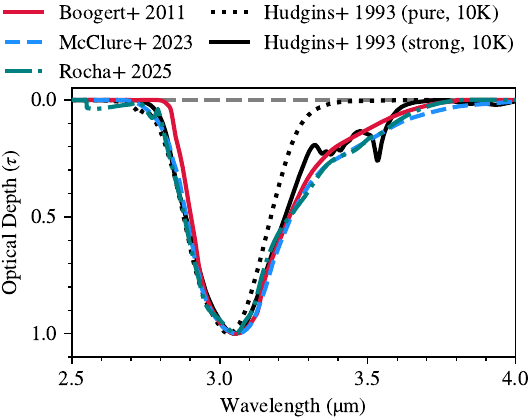}}
        \caption[]{Normalized absorption profiles of the \SI{3}{\micro\meter} feature toward two background stars published by \citet{Boogert2011,McClure2023} and one protostar \citep{Rocha2025}, compared with laboratory ice spectra from \citet{Hudgins1993}. The observed profiles (red, blue, and green) were extracted from published data and interpolated onto a common wavelength grid. Laboratory spectra correspond to pure water ice (dotted black) and a strong ice mixture (solid black) at \SI{10}{K}. All profiles are normalized to a peak optical depth of 1 for a direct comparison of the profile shape.}
    \label{fig:ice_profiles}
\end{figure}

Any variation in the shape of the \SI{3}{\micro\meter} ice absorption profile will naturally be reflected in the measured color excess. Several studies have reported that the detailed structure of this absorption feature can vary across different environments \citep[see, e.g.,][]{Boogert2015}. In particular, the composition of the ice mixture has a pronounced effect on the red wing of the profile, with species such as NH\(_3\)\(\cdot\)H\(_2\)O and CH\(_3\)OH contributing significantly \citep[e.g.,][]{Hudgins1993,Chiar2011,McClure2023,Rocha2025}. More recently, \citet{Noble2024} used \textit{JWST} observations to reveal minor contributions from so-called dangling OH groups on the blue side of the feature, near \SI{2.7}{\micro\meter}.

Despite these variations, there is also evidence that the overall shape of the \SI{3}{\micro\meter} absorption profile toward background stars is relatively stable \citep{Boogert2011,Madden2022}. To further investigate this, I present in Fig.~\ref{fig:ice_profiles} a comparison of three measured absorption profiles, together with selected laboratory spectra from \citet{Hudgins1993}. The red profile corresponds to the star 2MASS J17112005-2727131, a background giant behind the B59 core in the Pipe Nebula. The blue line shows the \textit{JWST} spectrum for NIR38 \citep{McClure2023}, a background giant behind the Chamaeleon I molecular cloud, while the green line represents Ced 110 IRS4 \citep[2MASS J11064638-7722287;][]{Rocha2025}, a binary protostellar system in Chamaeleon I. Because the optical depth profiles for these sources are not publicly available, I used WebPlotDigitizer\footnote{\href{https://automeris.io}{https://automeris.io}} to extract the data from published figures, manually tracing each spectrum and interpolating the results onto a regular wavelength grid with minimal smoothing. Any gaps in the measured spectra were interpolated, and all curves were renormalized to a peak optical depth of 1. For the laboratory data, I selected two mixtures: the dotted black line shows pure water ice (\SI{100}{\percent} H\(_2\)O), while the solid black line represents a strong ice mixture (H\(_2\)O:CH\(_3\)OH:CO:NH\(_3\) = 100:10:1:1), both at a temperature of \SI{10}{K}.

\begin{figure}[t]
        \centering
        \resizebox{1.0\hsize}{!}{\includegraphics[]{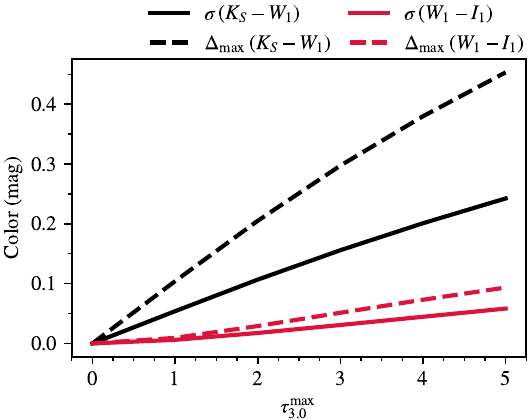}}
        \caption[]{Standard deviation (solid lines) and maximum difference (dashed lines) in \(K_S-W_1\) (black) and \(W_1-I_1\) (red) colors as a function of the peak optical depth of the \SI{3}{\micro\meter} feature, calculated for five absorption profiles (see Fig.~\ref{fig:ice_profiles}). The \(K_S-W_1\) color is significantly more sensitive to variations in the absorption profile, while \(W_1-I_1\) remains comparatively stable, enabling robust measurements of \taumax.}
    \label{fig:ice_profile_effect}
\end{figure}

Figure~\ref{fig:ice_profiles} reveals an apparent similarity among the observed absorption profiles, which also closely match the laboratory spectrum for the strong ice mixture. To quantify how variations in the absorption profile affect the measured color excess, I simulated the impact of each profile on broadband photometry. Specifically, I calculated the fluxes through different passbands as a function of the peak optical depth and derived the resulting color indices. The results are presented in Fig.~\ref{fig:ice_profile_effect}, which shows the standard deviation (solid lines) and maximum difference (dashed lines) in measured colors for all five absorption profiles as a function of optical depth. The black curves correspond to the \(K_S-W_1\) color, while the red curves represent \(W_1-I_1\).

These simulations demonstrate that the \(K_S-W_1\) color is much more sensitive to variations in the absorption profile than \(W_1-I_1\). This is because only the \(W_1\) band is affected by ice extinction in the former case, while the \(K_S\) band remains unaffected. In contrast, both \(W_1\) and \(I_1\) are affected in the latter, leading to a partial cancellation of profile-dependent effects. Even at large optical depths, the standard deviation in \(W_1-I_1\) remains at only a few percent. If the pure ice mixture is excluded from this test, the variation in both color indices decreases further: for a peak optical depth of 5, the standard deviation in \(K_S-W_1\) drops to about \SI{10}{\percent}, and in \(W_1-I_1\) to approximately \SI{3}{\percent}.

These findings demonstrate that, depending on the chosen color metric, the ice color excess method can be tuned to either enhance sensitivity to profile shape -- potentially enabling the detection of differences in ice composition -- or to minimize such sensitivity, thereby providing a robust measurement of \taumax.

\section{Selection of optimal passbands}
\label{sec:choice_passbands}

The selection of suitable photometric passbands is a critical step in reliably measuring \taumax. The broad \SI{3}{\micro\meter} ice feature, spanning roughly from \SI{2.7}{\micro\meter} to \SI{3.7}{\micro\meter}, is sampled by a limited set of widely available filters. As shown in Fig.~\ref{fig:spectra_passbands}, any practical combination for computing the ice color excess will necessarily include either \(W_1\), \(I_1\), or, in rare cases, \(L'\). Among these, \(L'\) only covers the red wing of the feature and does not sample the peak of the absorption feature near \SI{3}{\micro\meter}. Moreover, \(L'\) photometry is not broadly available and is typically restricted to small observing programs with specific targets. While such data may still be valuable for probing the red wing of the feature, large-scale studies must rely on the more widely available \(W_1\) and \(I_1\) passbands.

The options for constructing reliable color excess measurements are further constrained by the properties of other available bands. Both \(W_2\) and \(I_2\) are affected by additional ice absorption features and are therefore unsuitable for any color combination described in Sect.~\ref{sec:ice}. The only realistic choices for pairing with \(W_1\) or \(I_1\) are near-infrared passbands such as \(K_S\), which are accessible through large-area surveys like 2MASS, VVV(X), and VISIONS. This leaves three main configurations for calculating the ice color excess: \(\Lambda(K_S - W_1)\), \(\Lambda(K_S - I_1)\), and \lambdaWISEminusIRAC, as well as their use in the dust-reddening-free index \(Q'\).

The optimal choice of passbands is governed by two main factors: (a) the systematic error budget -- dominated by uncertainties in the extinction law -- and (b) the availability and quality of the photometric data. Pairing \(K_S\) with \(W_1\) offers the advantage of all-sky coverage, but is limited by the sensitivity of 2MASS and is subject to large systematic uncertainties due to the significant wavelength separation, which amplifies errors related to the extinction law. The same applies to \(K_S\) with \(I_1\), with the added limitation that \(I_1\) data do not cover the entire sky. The alternative approach, involving \(K_S\), \(W_1\), and \(I_1\) for \(Q'\), also suffers from large systematic uncertainties for the same reason.

\begin{figure}[t]
        \centering
        \resizebox{1.0\hsize}{!}{\includegraphics[]{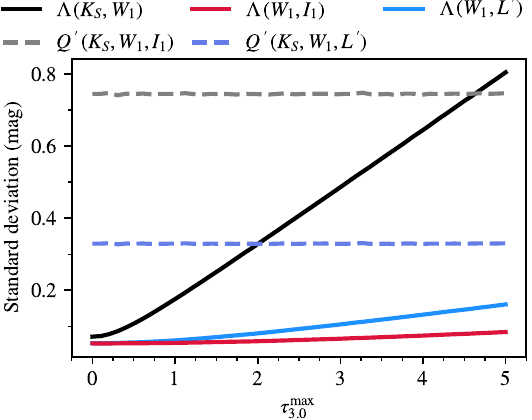}}
        \caption[]{Standard deviation of \(\Lambda\) and \(Q'\) color excess measurements as a function of the peak optical depth of the \SI{3}{\micro\meter} feature, derived from randomly sampling uncertainties in the photometric parameters and the extinction law. Solid lines represent \(\Lambda\) for different band combinations: \(\Lambda (K_S - W_1)\) in black, \lambdaWISEminusIRAC in red, and \(\Lambda (W_1 - L')\) in blue. Dashed lines show \(Q'\) for the combinations \(Q' (K_S, W_1, I_1)\) (gray) and \(Q' (K_S, W_1, L')\) (dark blue). The simulation demonstrates that \lambdaWISEminusIRAC and \(\Lambda(W_1 - L')\) maintain the lowest total uncertainties, even at high optical depths, due to the small wavelength separation between the bands.}
    \label{fig:errors_metrics}
\end{figure}

To quantify how the choice of passbands and the underlying uncertainties affect measurements of \(\Lambda\) and \(Q'\), I performed a Monte Carlo simulation that incorporates the various error sources discussed above. For each realization, random errors were assigned to the observed stellar magnitudes (\SI{0.01}{mag}), intrinsic stellar color (\SI{0.05}{mag}), and the measured dust extinction in the \(K_S\) passband (\SI{0.1}{mag}). Additionally, one of the seven extinction laws discussed above was randomly selected for each trial. Furthermore, a linear relation between extinction and the peak optical depth of the \SI{3}{\micro\meter} feature was assumed (\(\taumax = 0.5 A_{K_S}\)), consistent with literature findings and omitting an explicit ice formation threshold \citep[e.g.,][]{Boogert2011,Boogert2013,Madden2022}. This process was repeated one million times for \taumax values ranging from 0 to 5 in steps of 0.05. The results of this experiment are presented in Fig.~\ref{fig:errors_metrics}, which displays the standard deviation in the measured ice color excess metrics as a function of \taumax for several band combinations. Specifically, the figure shows \(\Lambda(K_S - W_1)\) (solid black line), \lambdaWISEminusIRAC (solid red line), \(\Lambda(W_1 - L')\) (solid blue line), \(Q'(K_S, W_1, I_1)\) (gray dashed line), and \(Q'(K_S, W_1, L')\) (dark blue dashed line).

Several key trends emerge from the simulation. Both \(Q'\) metrics exhibit relatively large but almost constant uncertainties across the full range of optical depths. This is because the error budget for \(Q'\) is dominated by uncertainties in the shape of the extinction law or, more specifically, the ratio of color excesses (see Eq.~\ref{equ:ice_alt}). In contrast, the various definitions of \(\Lambda\) generally yield smaller errors, with the uncertainty increasing nearly linearly with optical depth. The magnitude of this error is strongly dependent on the choice of band combination: \(\Lambda(K_S - W_1)\) shows large uncertainties, reflecting the greater impact of extinction law uncertainties when the wavelength separation between bands is large. Most notably, \lambdaWISEminusIRAC and \(\Lambda(W_1 - L')\) stand out for their remarkably small total error budgets, even at high optical depths. This is a direct consequence of the small wavelength separation between \(W_1\) and \(I_1\) (or \(L'\)).

These findings demonstrate that -- when considering both the error budget and data availability -- the combination of \(W_1\) and \(I_1\) in measuring \lambdaWISEminusIRAC provides the most robust and reliable measurement of the ice color excess. The results of the above described simulation highlight that systematic errors remain well controlled for this passband pair, even at high optical depths, due to the small wavelength separation and the resulting reduced sensitivity to extinction law uncertainties. Furthermore, both \(W_1\) and \(I_1\) offer high-precision photometry in key nearby star-forming regions, where future applications of this method are particularly effective thanks to the abundance of background stars and minimal confusion from overlapping clouds along the line of sight. For these reasons, the following analysis focuses on the application of the \ice using the \(W_1-I_1\) color:
\begin{equation}
\label{equ:ice_w1i1}
\Lambda\left(W_1 - I_1\right) = \left( W_1 - I_1 \right) - \left( W_1 - I_1 \right)_0 - A_{K_S} \cdot \Psi\left(W_1, I_1\right).
\end{equation}
Future studies may extend this approach to other band combinations as our understanding of the dust extinction law improves.

\section{Data and observational resources}
\label{sec:data}

To calibrate a parametrization for \taumax as a function of \lambdaWISEminusIRAC, I utilize multiple data sources to obtain infrared photometry and spectroscopically measured optical depths. This section provides an overview of the data archives employed and the specific quality criteria used for source selection.

\subsection{Infrared photometry}
\label{sec:infrared_photometry}

Infrared broadband photometry covering the \SI{3}{\micro\meter} absorption feature is the basis for the computation of the color excess caused by icy grains. In this wavelength range, two major data sources are available, each providing access to millions of individual measurements: data collected by the \textit{Spitzer} Space Telescope and data from the \textit{WISE} allsky survey. Both missions provide data for a set of passbands, including the filters \(W_1\) and \(I_1\) that nearly completely cover the broad ice absorption feature at \SI{3}{\micro\meter} (see Fig.~\ref{fig:spectra_passbands}). 

Both \textit{WISE} and \textit{Spitzer} provide multiple data releases, including versions tailored to specific sub-surveys conducted by the astronomical community, each of which may itself have several data releases. For \textit{WISE}, in this manuscript I use the unWISE catalog \citep{Schlafly2019}, which is based on coadds of all publicly available \textit{WISE} images in the \(W_1\) and \(W_2\) passbands. The unWISE data processing methodology improves the sensitivity limit by approximately \SI{0.7}{mag} compared to the AllWISE data release \citep{Cutri2014}.

For \textit{Spitzer} data, several catalogs are accessible via the NASA/IPAC Infrared Science Archive\footnote{\href{https://irsa.ipac.caltech.edu/}{https://irsa.ipac.caltech.edu/}} (IRSA). Among these, the most relevant are those covering prominent, nearby, and actively star-forming molecular clouds, such as Orion, Taurus, Perseus, Ophiuchus, Lupus, Chamaeleon, and Corona Australis. However, data processing methods -- and thus data quality -- vary across \textit{Spitzer} products due to the use of diverse pipelines and reduction techniques. To mitigate potential systematic differences, I adopt the Spitzer Extended Solar Neighborhood Archive (SESNA; see \citealp{Pokhrel2020} for an overview and a forthcoming publication by Gutermuth et al. for details) as the primary source of \textit{Spitzer} photometry. SESNA provides a uniformly reduced catalog for several dozen nearby molecular cloud complexes observed during the operational lifetime of \textit{Spitzer}. 

Some regions, however, are not included in SESNA. Notably, the Taurus star-forming complex \citep{Rebull2010} has not yet been reprocessed in the context of the SESNA project, but catalog data for this region are available through IRSA \citep{Taurus2020}. For all remaining lines of sight not covered by SESNA or the Taurus data release, I use data from the \textit{Spitzer} survey "From Molecular Cores to Planet-Forming Disks" \citep[c2d;][]{Evans2003}.

Near-infrared (NIR) \(K_S\)-band photometry is sourced exclusively from the Two Micron All Sky Survey \citep{Skrutskie2006}. For future applications of the technique presented in this manuscript, more recent wide-field NIR surveys -- such as the VISTA Variables in the Via Láctea \citep[VVV;][]{Minniti2010}, its extension VVVX, and especially the VISTA Star Formation Atlas \citep[VISIONS;][]{Meingast2023} -- will be of increasing relevance. The VVV and VVVX surveys cover the Galactic bulge and approximately \SI{150}{deg} in Galactic longitude along the Milky Way plane. In contrast, VISIONS targets five nearby (d \(\lesssim\) \SI{500}{pc}) star-forming regions, providing ideal coverage for future measurements of the ice color excess.

\subsection{Optical depth sample}
\label{sec:optical_depth_data}

\begin{table}[t!]
\caption{Overview of the literature sources used for the calibration sample, listing the number of sources included in the final selection (and the total number published), the observing facilities, and the symbols used to represent each dataset in subsequent figures.
}
\label{tab:literature_overview}
\centering
\begin{tabular}{@{\extracolsep{\fill}} l c c c c}
\hline\hline
Reference & N (total) & Facility & Symbol \\
\hline
\citet{Murakawa2000}   & 4 (61)     & WIRO/MLOF & \tikz[baseline=-0.6ex]\draw[draw=edgegray, line width=0.3mm, fill=purple] (0,0)--(0.09,0.12)--(0.18,0)--(0.09,-0.12)--cycle; \\
\citet{Boogert2011}    & 17 (33)  & Spitzer & \tikz\draw[draw=edgegray, line width=0.3mm, fill=red] (0,0) circle (0.11); \\
\citet{Chiar2011}      & 5 (10)   & IRTF & \tikz\draw[draw=edgegray, line width=0.3mm, fill=yellow] (0,0.12)--(0.12,0)--(0,-0.12)--(-0.12,0)--cycle; \\
\citet{Boogert2013}    & 12 (32)  & Spitzer & \tikz\draw[draw=edgegray, line width=0.3mm, fill=brightgreen] (90:0.13)--(162:0.13)--(234:0.13)--(306:0.13)--(18:0.13)--cycle; \\
\citet{Noble2013}      & 0 (30)   & AKARI & \\
\citet{Goto2018}       & 1 (21)   & IRTF & \tikz\draw[draw=edgegray, line width=0.3mm, fill=green] (0,0)--(-0.2,0.12)--(-0.2,-0.12)--cycle; \\
\citet{Madden2022}     & 17 (49)  & Spitzer & \tikz\draw[fill=lightblue] (0,0) rectangle (0.2,0.2); \\
\citet{McClure2023}    & 0 (2)    & JWST & \\
\citet{Smith2025}      & 0 (44)   & JWST & \\
\hline
\end{tabular}
\end{table}

\begin{figure*}[t]
        \centering
        \resizebox{1.0\hsize}{!}{\includegraphics[]{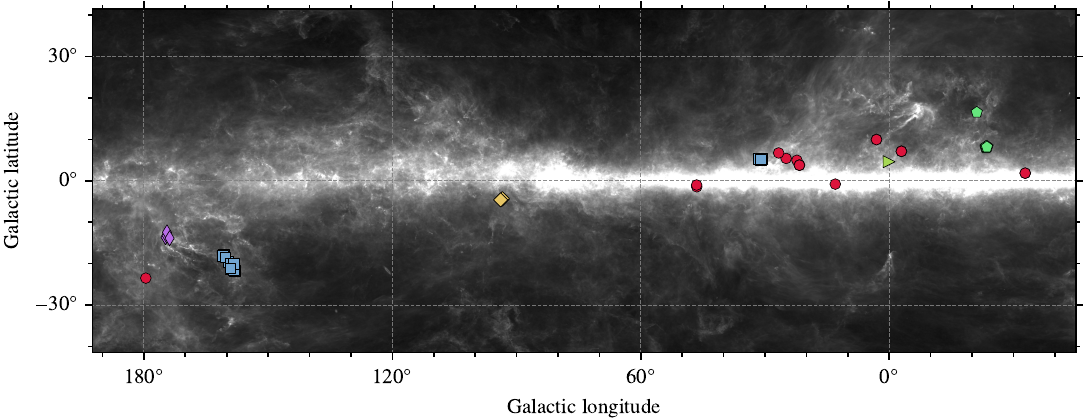}}
        \caption[]{Positions of the final literature sample of background stars with spectroscopically measured \SI{3}{\micro\meter} optical depths, overlaid on a \textit{Planck} \SI{857}{\giga\hertz} (\SI{350}{\micro\meter}) image of the Milky Way. Different symbols (colors and shapes) indicate the data source, as detailed in Table~\ref{tab:literature_overview}. The sample spans more than \SI{180}{\deg} in Galactic longitude, covering a diverse set of nearby star-forming regions.}
    \label{fig:sources_allsky}
\end{figure*}

To construct a robust parametrization of \taumax, I conducted an extensive literature search for spectroscopic measurements of this feature toward stars in the background of molecular clouds. This search identified nine major sources, encompassing a broad range of observing facilities, data quality, and target environments. A summary of the literature sources, telescope facilities, and corresponding symbols used in subsequent plots is provided in Table~\ref{tab:literature_overview}. A detailed overview of the final source list, including adopted spectral types, photometric parameters, extinction, and measured ice color excesses, is given in Table~\ref{tab:master}.

\citet{Murakawa2000} used the Wyoming Infrared Observatory (WIRO)\footnote{\href{https://physics.uwyo.edu/~WIRO/}{https://physics.uwyo.edu/~WIRO/}} and the Mt. Lemmon Observing Facility (MLFO)\footnote{\href{https://cse.umn.edu/mifa/research/lemmon}{https://cse.umn.edu/mifa/research/lemmon}} to observe 61 background stars in the Taurus Molecular Cloud. Out of the 61 observed sources, the authors report \taumax values for 55 sources, including many cases where the feature was not detected (\(\taumax = 0\)). \citet{Boogert2011} combined \textit{Spitzer} IRS spectra with ground-based Keck data to observe 33 sources across regions such as Ophiuchus, Serpens, Aquila, and Taurus, thereby sampling a variety of astrophysical environments. For two of these sources, only upper limits of \taumax were reported. \citet{Chiar2011} used IRTF and \textit{Spitzer}/IRS to obtain spectra for ten sources in the IC 5146 (Barnard 168) cloud, with the SpeX instrument \citep{Rayner2003}. Of these, for seven sources \taumax was measured. \citet{Boogert2013} presented \textit{Spitzer}/IRS observations of 32 sources in the Lupus cloud complex. Of these, for 11 sources the \taumax measurement corresponds to an upper limit. \citet{Noble2013} provided 30 \textit{AKARI} \citep{Onaka2007,Ohyama2007} spectra, covering regions such as Ophiuchus, Lupus, and Taurus. In this publication, explicit values for \taumax were not reported. Hence, I estimated values by eye from the published plots for 13 sources with unambiguous features. \citet{Goto2018} used IRTF/SpeX to observe 21 sources, providing usable \taumax values for seven sources in the Pipe Nebula, with the remaining measurements being upper limits. \citet{Madden2022} recorded 49 spectra projected against Perseus and Serpens using \textit{Spitzer}/IRS and IRTF/SpeX, with 28 \taumax measurements in Perseus and 21 in Serpens. In more recent work, \citet{McClure2023} used \textit{JWST} to observe two highly obscured sources behind the Chamaeleon I cloud, and \citet{Smith2025} expanded this with \textit{JWST} spectra for 44 background sources in the same region.

For all data sources, I applied a consistent workflow. First, I extracted coordinates, \taumax (excluding upper limits), and extinction measurements from the original publications, converting all coordinates to the ICRS reference frame as needed. For the earliest published data from \citet{Murakawa2000}, for which coordinates were published in J1950, I visually inspected 2MASS images using Aladin and manually corrected mismatches resulting from imperfect coordinate transformations. For all literature sources reporting individual extinction values in the \(V\) passband, these were converted to \(A_{K_S}\) using the extinction law of \citet{Boogert2011}. Next, I cross-matched the compiled catalogs with the above-listed infrared datasets, including unWISE, c2d, SESNA, and \textit{Spitzer} Taurus, using a matching radius of \SI{2}{arcsec}. In all cases, unWISE source coordinates were used as the reference. For \textit{Spitzer} photometry, I preferred SESNA data, using Taurus and c2d as fallback options in this order.

To ensure high data quality, I imposed a series of strict selection criteria. Foremost, I required an unambiguous cross-match with the unWISE catalog. Additional requirements included an unWISE quality factor on \(W_1\) (\texttt{qfW1}) greater than 0.99, a \(W_1\) magnitude error less than \SI{2}{mmag}, unWISE Coadd Flags at the central pixel less than or equal to 1, and a \(W_1\) inter-quartile range from unTimely time series below \SI{0.04}{mag}. For sources with \(I_1\) data from c2d, I required IR1\_Q\_DET\_C = A, IR1\_Q\_FLUX\_M = A, and \(10 < \text{IR1\_FLUX\_C} < 100\). Moreover, for all sources, the spectral type had to be listed in the corresponding reference.

After cross-matching and applying these quality criteria, the initial set of 282 unique sources was reduced to a final sample of 56 sources suitable for analysis. Notably, none of the \textit{JWST} measurements could be included, primarily because these sources are too faint to be reliably detected by both \textit{Spitzer} or \textit{WISE}. Specifically, for the prominent source NIR38 first investigated by \citet{McClure2023} both \textit{Spitzer} and \textit{WISE} measurements are available, but the source was excluded from subsequent analysis due to significant photometric variability (unWISE peak-to-peak amplitude of \SI{0.31}{mag}). Conversely, many targets from the earlier publications were removed for being too bright for these satellites. In addition, none of the \textit{AKARI} measurements from \citet{Noble2013} passed the quality cuts, primarily due to missing quality flags for the flux measurements in the c2d catalog.

Figure~\ref{fig:sources_allsky} shows the positions of the final source selection overlaid on a view of the Milky Way as observed by \textit{Planck} \citep{PlanckI} at \SI{857}{\giga\hertz} (\SI{350}{\micro\meter}). Different symbols (colors and shapes) represent different literature sources as listed in Table~\ref{tab:literature_overview}. The selected sources span galactic longitudes over more than \SI{180}{deg}, sampling a wide variety of nearby star-forming regions and providing representative coverage of astrophysical conditions.

\section{Empirical calibration}
\label{sec:calibration}

With the mathematical framework, the choice of passbands, and the selection of calibration sources from the literature in place, in this section, I present the empirical parametrization for fitting the peak optical depth of the \SI{3}{\micro\meter} ice feature, \taumax, as a function of the ice color excess \lambdaWISEminusIRAC. Since the relationship between \taumax and any passband combination for \(\Lambda\) is generally non-linear, I adopt a function of the form
\begin{equation}
\label{equ:ice_fit_func}
\taumax = a \cdot \ln(1 + b \cdot \Lambda),
\end{equation}
with \(a\) and \(b\) being the free parameters. Furthermore, the form of this equation is physically motivated to pass through the origin: an ice color excess of \(\Lambda = 0\) must correspond to zero optical depth.

To accurately represent measurement errors in both \(\Lambda\) and \taumax, several factors are considered. The error on \taumax is not always reported in the literature; when missing, I assume a standard deviation of \SI{10}{\percent} of the value itself. The error budget for \(\Lambda\) is more complex (see Sect.\ref{sec:errors}). Photometric errors are taken directly from the source catalogs (unWISE, SESNA, \textit{Spitzer} Taurus, or c2d). For intrinsic colors, I use the smoothed fit from Fig.~\ref{fig:intrinsic_colors} and adopt a fixed error of \SI{0.02}{mag}. Errors on extinction (\(A_{K_S}\)) are taken from the literature when available, or otherwise set to \SI{0.1}{mag}. Additionally, I include a systematic error term to account for stellar variability, using the inter-quartile range from unWISE time series photometry, which is typically \SI{0.01}{mag}. All values for \taumax and \lambdaWISEminusIRAC used for the fit, and their associated uncertainties are listed in Table~\ref{tab:master} for the final source selection.

The choice of the specific extinction law is a critical aspect of the calibration. Since most high-optical-depth sources in the final sample (\(\taumax \gtrsim 0.5\)) are from \citet{Boogert2011}, I adopt the extinction law published in that work. As discussed in Sect.~\ref{sec:errors_extinction_law}, using \lambdaWISEminusIRAC minimizes sensitivity to the extinction law, but I emphasize that this calibration is only strictly valid when used with this specific law; caution is required when applying it to other datasets.

\begin{figure}[t]
        \centering
        \resizebox{1.0\hsize}{!}{\includegraphics[]{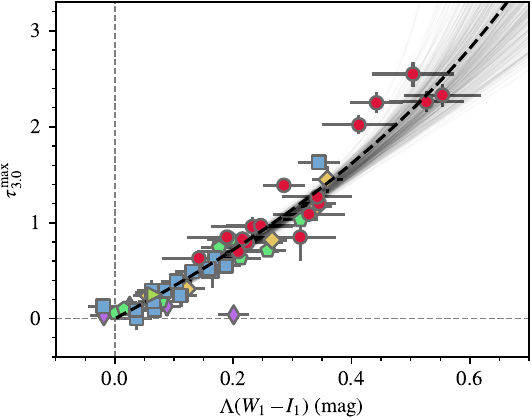}}
        \caption[]{Peak optical depth of the \SI{3}{\micro\meter} ice absorption feature (\taumax) as a function of the ice color excess \lambdaWISEminusIRAC for the calibration sample. The various symbols represent the different literature sources listed in Table~\ref{tab:literature_overview}. Each point represents a literature measurement with propagated uncertainties. The black dashed line shows the best-fit relation, while thin solid lines represent \num{500} random samples from the fit parameter space. Gray dashed lines mark \(\taumax = 0\) and \(\lambdaWISEminusIRAC = 0\).}
    \label{fig:ice_literature}
\end{figure}

To fit the data and account for measurement errors in both variables, I perform the fit using the SciPy \texttt{curve\_fit} function, repeating the procedure \num{100000} times. In each iteration, I randomly sample all input data (\taumax and \lambdaWISEminusIRAC) from normal distributions defined by the calculated mean value and the propagated uncertainties. This yields a distribution of possible fits and allows for robust estimation of uncertainties in both the \(a\) and \(b\) parameters in Eq.\ref{equ:ice_fit_func}. Given the non-linear parameter space, I report the geometric median as the best-fit solution, i.e., the point minimizing the sum of Euclidean distances to all sampled points. The resulting fit is shown in Fig.~\ref{fig:ice_literature}, where individual data points correspond to literature measurements, the dashed black line shows the best-fit relation, and gray thin solid lines represent \num{500} random samples from the parameter space. Gray dashed lines mark \(\taumax = 0\) and \(\lambdaWISEminusIRAC = 0\), the physically meaningful boundaries. The non-linear distribution of \(a\) and \(b\) values from the sampling is shown in Fig.~\ref{fig:ice_fit_error}.

Figure~\ref{fig:ice_literature} reveals a strikingly tight relation between the peak optical depth \taumax and the computed ice color excess \lambdaWISEminusIRAC across the entire range of presented optical depths. Only one clear outlier is present, corresponding to a source from \citet{Murakawa2000} (source 14 in Table~\ref{tab:master}), which is likely affected by contamination from a nearby bright star. All other measurements fall along a well-constrained relation, resulting in a robust fit. The best-fitting solution is 
\begin{equation}
\label{equ:ice_fit_final}
    \taumax = -3.51 \cdot \ln\left[1 - 0.92 \cdot \lambdaWISEminusIRAC\right].
\end{equation}
This tight correlation is all the more remarkable given that the data are drawn from a diverse set of sources, each often relying on different telescope facilities, instrumentation, and data processing pipelines, including the methods used to estimate line-of-sight extinction, approaches to continuum fitting, and techniques for measuring \taumax. The consistency observed here serves as a strong testament to the robustness and general applicability of the ice color excess method.

For error propagation in the derived equation, it is advisable to sample the parameters \(a\) and \(b\) in logarithmic space, given the non-linearity of the parameter space; specifically, \(\ln(-a) = 1.36 \pm 0.5\) and \(\ln(-b) = -0.21 \pm 0.43\). These large 1-\(\sigma\) uncertainties in the logarithmic fit parameters primarily arise from the degeneracy between \(a\) and \(b\) in the fit function, rather than from scatter in the input data. This degeneracy allows a broad range of parameter combinations to provide acceptable fits, as illustrated in Fig.~\ref{fig:ice_fit_error}. Note that these quoted mean values are intended solely for sampling and do not correspond to the reported best-fit values (i.e., the geometric median).

\begin{figure}[t]
        \centering
        \resizebox{1.0\hsize}{!}{\includegraphics[]{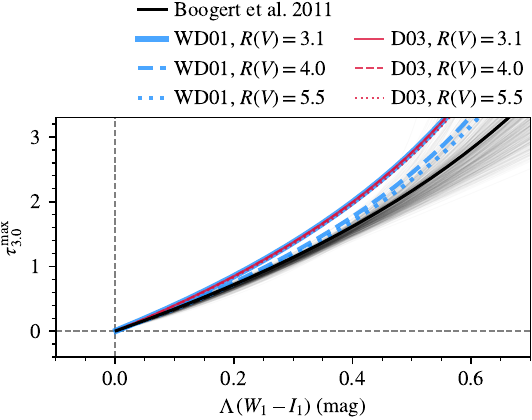}}
        \caption[]{Relation between the peak optical depth of the \SI{3.0}{\micro\meter} ice feature (\taumax) and the ice color excess \lambdaWISEminusIRAC for different assumptions about the extinction law. The fitted curves correspond to extinction laws from \citet{Weingartner2001} (blue), \citet{Draine2003} (red), and \citet{Boogert2011} (black). While the use of the \lambdaWISEminusIRAC metric minimizes the impact of extinction law uncertainties, systematic differences between the fits remain visible. However, these differences are generally within the typical measurement uncertainties of the calibration. This highlights the importance of adopting a consistent extinction law or recalibrating the method when applying it to different environments.}
    \label{fig:ice_literature_alt_extinction_law}
\end{figure}

In Sect.~\ref{sec:errors_extinction_law}, I extensively discussed the systematic effect of the shape of the extinction law on the applicability of the method. For reference, I present in Fig.~\ref{fig:ice_literature_alt_extinction_law} the fits for the relation between \taumax and \lambdaWISEminusIRAC, when different extinction laws are used for the determination of the factor \(\Psi\) in Eq.~\ref{equ:ice_w1i1}. The considered extinction laws are those published by \citet{Weingartner2001} (blue lines), \citet{Draine2003} (red lines), and \citet{Boogert2011} (black line). The thin black lines correspond to \num{500} random samples of the fit with the \citet{Boogert2011} extinction law.

The figure demonstrates that even in the case of choosing \lambdaWISEminusIRAC, which minimizes the impact of the extinction law, significant differences between the fits become visible. However, in the case of using \lambdaWISEminusIRAC, the presented curves are still within the typical range of uncertainties in the adopted fit of Eq.~\ref{equ:ice_fit_final}.

\section{Discussion}
\label{sec:discussion}

The \ice technique presented in this work demonstrates considerable promise for mapping and quantifying the distribution of interstellar ices using widely available infrared broadband photometry. In the formulation presented in this manuscript, the calibration of the method benefits from well-characterized samples where the intrinsic colors of background stars are known -- typically through accurate spectral typing. Moreover, for the individual sources, the extinction along the line of sight has been independently determined from spectroscopic measurements. This controlled scenario allows for a robust calibration of the ice color excess and the parameterization of \taumax. However, scaling this approach to larger and more diverse samples introduces new challenges. In particular, variations in intrinsic stellar colors and uncertainties in extinction estimates become increasingly important when dealing with heterogeneous stellar populations. Future applications will need to address these issues, for example by employing probabilistic methods for stellar classification or leveraging \textit{Gaia} parallaxes and multi-band photometry to better constrain stellar parameters and extinction.

In particular, the ability to generate spatially resolved ice maps at (sub-)arcminute scales in nearby star-forming regions represents a significant advance over existing techniques, which are often limited by the depth or coverage of spectroscopic surveys. The prospects for applying this technique on a Galactic scale are exceptionally promising. The SESNA catalog alone contains over eight million sources with high-quality infrared photometry, many of which have direct counterparts in the \textit{WISE} all-sky survey. Even more, the \textit{Spitzer} program Galactic Legacy Infrared Midplane Survey Extraordinaire \citep[GLIMPSE;][]{Benjamin2003} contains about 50 million sources, enabling ice mapping in the Galactic Plane. However, similar to dust extinction mapping techniques, multiple overlapping clouds along the line of sight create another challenge to address in constructing \taumax maps \citep{Zhang2022}. Nevertheless, applying the presented method on such scales opens the door to constructing high-resolution, wide-area maps of \taumax. Such maps would enable detailed studies of the relationship between ice formation, dust properties, and the local interstellar environment. 

Moreover, expanding and recalibrating the method to incorporate near-infrared data from 2MASS, VVV(X), and VISIONS increases the number of usable sources by an order of magnitude. In this regard, particularly the 2MASS and unWISE datasets enable an all-sky application of the method. However, a key challenge in extending the \ice method to NIR photometry is the selection of the extinction law, as the conversion from color excess to \taumax is highly sensitive to the adopted extinction curve. The extinction law is known to vary significantly, even on relatively small spatial scales \citep[e.g.,][]{Green2024,Zhang2025}, and the \ice method is especially sensitive when combining \(K_S\)-band photometry with \(W_1\) data (see Sect.\ref{sec:choice_passbands}). Therefore, it is crucial to employ a standardized approach and, where feasible, recalibrate the method to account for different extinction law characteristics or at least incorporate this systematic effect into the error estimation. The compiled source catalog provided in this work (Table\ref{tab:master}) can act as a valuable resource for such recalibrations, enabling users to tailor the technique to their specific datasets and scientific objectives.


Further, it is instructive to compare the capabilities of this photometric approach with those of \textit{JWST}, which has rapidly established itself as a premier facility for ice chemistry studies. \textit{JWST}'s
unparalleled sensitivity and spectral resolution in the mid-infrared have revealed exquisite detail in the absorption features of interstellar ices, uncovering subtle compositional variations and previously unknown molecular species \citep[e.g.,][]{McClure2023,Dartois2024,Rocha2025}. While the \ice method cannot provide the same level of chemical detail or resolve fine spectral structure, it serves as a powerful tool for parameterizing the peak optical depth of the main ice features across large samples and broad spatial scales. In this way, the technique complements the detailed, small-area studies enabled by JWST \citep{Smith2025}, allowing to place localized chemical findings into a global Galactic context. Photometric ice maps can be used to investigate ice formation thresholds as reported by, for example, \citet{Chiar2011,Boogert2011,Boogert2013,Madden2022}. However, the sample sizes in these studies have been too limited to report these findings in the context of the greater galactic environment and to investigate the relation of ice properties to, for instance, the star-formation history of local cloud complexes \citep[e.g.,][]{Posch2023,Ratzenboebck2023,Swiggum2024} and the impact of OB associations on nearby dust clouds \citep[e.g.,][]{Alves2025}. For instance, with the presented method it will become possible to correlate ice abundance with the radiative influence of massive stars and the dust temperature distributions measured by the \textit{Planck} and \textit{Herschel} \citep{Herschel} missions and published by, for example \citet{Lombardi2014}, \citet{Zari2016}, \citet{Lada2017}, and \citet{Hasenberger2018}, or those available in the \textit{Herschel} Gould Belt Survey Archive\footnote{\href{http://www.herschel.fr/cea/gouldbelt/en/index.php}{http://www.herschel.fr/cea/gouldbelt/en/index.php}}.



\section{Summary}
\label{sec:summary}

This study presents the \ice method, a new, scalable technique for quantifying the peak optical depth of the \SI{3}{\micro\meter} ice absorption feature (\taumax). The primary goal was to develop and empirically calibrate this procedure, which leverages data from publicly available photometric surveys. The method is based on the concept of the ice color excess, which quantifies the amount of reddening caused by absorption features of icy dust grains, and is calibrated using a carefully selected sample of literature sources with spectroscopically measured optical depths. The main findings of this work are:
\begin{itemize}
    \item I developed two techniques for parameterizing \taumax: (a) the color excess method (index \(\Lambda\); \hyperref[sec:definition]{Sect.2.2}), and (b) the reddening-free index method (index \(Q'\); \hyperref[sec:alt_definition]{Sect.2.3}). The practical application of these quantities is formalized in Eq.~\eqref{equ:ice} for the color excess approach, and in Eq.~\eqref{equ:ice_alt} for the reddening-free index approach. The indices \(\Lambda\) and \(Q'\) can be constructed with a diverse set of infrared passbands.
    \item All approaches, and especially the selection of passbands, require careful consideration of systematic uncertainties, including the adopted extinction law, intrinsic stellar colors, source variability, and the optical depth profile of the ice absorption feature (Sect.~\ref{sec:errors}). Among these, the choice of extinction law (Sect.~\ref{sec:errors_extinction_law}) is the most significant source of systematic error.
    \item Through a detailed analysis of error propagation and data availability (Sect.~\ref{sec:choice_passbands}), I identified the combination of the \textit{WISE} \(W_1\) and \textit{Spitzer} \(I_1\) passbands as the optimal metric to measure \(\Lambda\) in the form of \(\lambdaWISEminusIRAC\). This choice minimizes systematic uncertainties due to the small wavelength separation of the passbands and leverages the high-quality and high-precision photometry available in the most relevant nearby star-forming regions. All suitable passband combinations for constructing \(Q'\) result in large uncertainties due to unknown specifics of the extinction law (Fig.~\ref{fig:errors_metrics}).
    \item I compiled a literature sample of spectroscopically measured optical depths. These sources largely used different telescopes, instruments, and data processing methods to measure \taumax. To search for counterparts in infrared data archives, I queried the unWISE, SESNA, c2d, and \textit{Spitzer} Taurus source catalogs (Sect.~\ref{sec:optical_depth_data}, Table~\ref{tab:literature_overview}). Further, I applied stringent quality criteria to ensure a robust calibration, leaving a total of 56 sources to fit \taumax as a function of \lambdaWISEminusIRAC.
    \item I present an empirical calibration of the ice color excess method, finding a remarkably tight and well-determined correlation between \taumax and \lambdaWISEminusIRAC (Sect.~\ref{sec:calibration}, Fig.~\ref{fig:ice_literature}). The best-fitting relation, using the \citet{Boogert2011} extinction law is given by Eq.~\eqref{equ:ice_fit_final}.
\end{itemize}
The \ice method developed here enables efficient, large-scale mapping of interstellar ices across the Milky Way. By combining archival photometric data with a robust empirical calibration, this approach enables mapping of the distribution and environmental dependence of the occurrence of icy dust grains in Galactic molecular clouds and star-forming regions.

\begin{acknowledgements}

I thank the anonymous referee for the useful comments that helped to improve this publication.

This work made use of Astropy:\footnote{\href{https://www.astropy.org}{https://www.astropy.org}} a community-developed core Python package and an ecosystem of tools and resources for astronomy \citep{astropy2013, astropy2018, astropy2022}.

This research has made use of the SIMBAD database operated at CDS, Strasbourg, France \citep{simbad}. 

This research has made use of the VizieR catalogue access tool, CDS, Strasbourg, France \citep{10.26093/cds/vizier}. The original description of the VizieR service was published in \citet{vizier2000}.

This research has made use of "Aladin sky atlas" developed at CDS, Strasbourg Observatory, France \citep{Bonnarel2000}.

This research made use of hips2fits,\footnote{\href{https://alasky.cds.unistra.fr/hips-image-services/hips2fits}{https://alasky.cds.unistra.fr/hips-image-services/hips2fits}} a service provided by CDS.

This research has made use of the SVO Filter Profile Service \citep{svo2012,svo2020}.

This research has made use of the NASA/IPAC Infrared Science Archive, which is funded by the National Aeronautics and Space Administration and operated by the California Institute of Technology.

This research has made use of the \textit{Planck} Legacy Archive\footnote{\href{https://pla.esac.esa.int/pla}{https://pla.esac.esa.int/pla}}.

I also acknowledge usage of other various Python packages in the data analysis of this work, including astroquery \citep{astroquery}, Matplotlib \citep{matplotlib}, NumPy \citep{numpy}, Pandas \citep{Pandas1,Pandas2}, scikit-learn \citep{scikit-learn}, and SciPy \citep{scipy}.

\end{acknowledgements}

\bibliography{references.bib}

\begin{thebibliography}{89}
\expandafter\ifx\csname natexlab\endcsname\relax\def\natexlab#1{#1}\fi

\bibitem[{{Alves} {et~al.}(2025){Alves}, {Lombardi}, \& {Lada}}]{Alves2025}
{Alves}, J., {Lombardi}, M., \& {Lada}, C. 2025, arXiv e-prints,
  arXiv:2501.13931

\bibitem[{{Astropy Collaboration} {et~al.}(2022){Astropy Collaboration},
  {Price-Whelan}, {Lim}, {Earl}, {Starkman}, {Bradley}, {Shupe}, {Patil},
  {Corrales}, {Brasseur}, {N{\"o}the}, {Donath}, {Tollerud}, {Morris},
  {Ginsburg}, {Vaher}, {Weaver}, {Tocknell}, {Jamieson}, {van Kerkwijk},
  {Robitaille}, {Merry}, {Bachetti}, {G{\"u}nther}, {Aldcroft},
  {Alvarado-Montes}, {Archibald}, {B{\'o}di}, {Bapat}, {Barentsen},
  {Baz{\'a}n}, {Biswas}, {Boquien}, {Burke}, {Cara}, {Cara}, {Conroy},
  {Conseil}, {Craig}, {Cross}, {Cruz}, {D'Eugenio}, {Dencheva}, {Devillepoix},
  {Dietrich}, {Eigenbrot}, {Erben}, {Ferreira}, {Foreman-Mackey}, {Fox},
  {Freij}, {Garg}, {Geda}, {Glattly}, {Gondhalekar}, {Gordon}, {Grant},
  {Greenfield}, {Groener}, {Guest}, {Gurovich}, {Handberg}, {Hart},
  {Hatfield-Dodds}, {Homeier}, {Hosseinzadeh}, {Jenness}, {Jones}, {Joseph},
  {Kalmbach}, {Karamehmetoglu}, {Ka{\l}uszy{\'n}ski}, {Kelley}, {Kern},
  {Kerzendorf}, {Koch}, {Kulumani}, {Lee}, {Ly}, {Ma}, {MacBride}, {Maljaars},
  {Muna}, {Murphy}, {Norman}, {O'Steen}, {Oman}, {Pacifici}, {Pascual},
  {Pascual-Granado}, {Patil}, {Perren}, {Pickering}, {Rastogi}, {Roulston},
  {Ryan}, {Rykoff}, {Sabater}, {Sakurikar}, {Salgado}, {Sanghi}, {Saunders},
  {Savchenko}, {Schwardt}, {Seifert-Eckert}, {Shih}, {Jain}, {Shukla}, {Sick},
  {Simpson}, {Singanamalla}, {Singer}, {Singhal}, {Sinha}, {Sip{\H{o}}cz},
  {Spitler}, {Stansby}, {Streicher}, {{\v{S}}umak}, {Swinbank}, {Taranu},
  {Tewary}, {Tremblay}, {de Val-Borro}, {Van Kooten}, {Vasovi{\'c}}, {Verma},
  {de Miranda Cardoso}, {Williams}, {Wilson}, {Winkel}, {Wood-Vasey}, {Xue},
  {Yoachim}, {Zhang}, {Zonca}, \& {Astropy Project Contributors}}]{astropy2022}
{Astropy Collaboration}, {Price-Whelan}, A.~M., {Lim}, P.~L., {et~al.} 2022,
  \apj, 935, 167

\bibitem[{{Astropy Collaboration} {et~al.}(2018){Astropy Collaboration},
  {Price-Whelan}, {Sip{\H{o}}cz}, {G{\"u}nther}, {Lim}, {Crawford}, {Conseil},
  {Shupe}, {Craig}, {Dencheva}, {Ginsburg}, {VanderPlas}, {Bradley},
  {P{\'e}rez-Su{\'a}rez}, {de Val-Borro}, {Aldcroft}, {Cruz}, {Robitaille},
  {Tollerud}, {Ardelean}, {Babej}, {Bach}, {Bachetti}, {Bakanov}, {Bamford},
  {Barentsen}, {Barmby}, {Baumbach}, {Berry}, {Biscani}, {Boquien}, {Bostroem},
  {Bouma}, {Brammer}, {Bray}, {Breytenbach}, {Buddelmeijer}, {Burke},
  {Calderone}, {Cano Rodr{\'\i}guez}, {Cara}, {Cardoso}, {Cheedella}, {Copin},
  {Corrales}, {Crichton}, {D'Avella}, {Deil}, {Depagne}, {Dietrich}, {Donath},
  {Droettboom}, {Earl}, {Erben}, {Fabbro}, {Ferreira}, {Finethy}, {Fox},
  {Garrison}, {Gibbons}, {Goldstein}, {Gommers}, {Greco}, {Greenfield},
  {Groener}, {Grollier}, {Hagen}, {Hirst}, {Homeier}, {Horton}, {Hosseinzadeh},
  {Hu}, {Hunkeler}, {Ivezi{\'c}}, {Jain}, {Jenness}, {Kanarek}, {Kendrew},
  {Kern}, {Kerzendorf}, {Khvalko}, {King}, {Kirkby}, {Kulkarni}, {Kumar},
  {Lee}, {Lenz}, {Littlefair}, {Ma}, {Macleod}, {Mastropietro}, {McCully},
  {Montagnac}, {Morris}, {Mueller}, {Mumford}, {Muna}, {Murphy}, {Nelson},
  {Nguyen}, {Ninan}, {N{\"o}the}, {Ogaz}, {Oh}, {Parejko}, {Parley}, {Pascual},
  {Patil}, {Patil}, {Plunkett}, {Prochaska}, {Rastogi}, {Reddy Janga},
  {Sabater}, {Sakurikar}, {Seifert}, {Sherbert}, {Sherwood-Taylor}, {Shih},
  {Sick}, {Silbiger}, {Singanamalla}, {Singer}, {Sladen}, {Sooley},
  {Sornarajah}, {Streicher}, {Teuben}, {Thomas}, {Tremblay}, {Turner},
  {Terr{\'o}n}, {van Kerkwijk}, {de la Vega}, {Watkins}, {Weaver}, {Whitmore},
  {Woillez}, {Zabalza}, \& {Astropy Contributors}}]{astropy2018}
{Astropy Collaboration}, {Price-Whelan}, A.~M., {Sip{\H{o}}cz}, B.~M., {et~al.}
  2018, \aj, 156, 123

\bibitem[{{Astropy Collaboration} {et~al.}(2013){Astropy Collaboration},
  {Robitaille}, {Tollerud}, {Greenfield}, {Droettboom}, {Bray}, {Aldcroft},
  {Davis}, {Ginsburg}, {Price-Whelan}, {Kerzendorf}, {Conley}, {Crighton},
  {Barbary}, {Muna}, {Ferguson}, {Grollier}, {Parikh}, {Nair}, {Unther},
  {Deil}, {Woillez}, {Conseil}, {Kramer}, {Turner}, {Singer}, {Fox}, {Weaver},
  {Zabalza}, {Edwards}, {Azalee Bostroem}, {Burke}, {Casey}, {Crawford},
  {Dencheva}, {Ely}, {Jenness}, {Labrie}, {Lim}, {Pierfederici}, {Pontzen},
  {Ptak}, {Refsdal}, {Servillat}, \& {Streicher}}]{astropy2013}
{Astropy Collaboration}, {Robitaille}, T.~P., {Tollerud}, E.~J., {et~al.} 2013,
  \aap, 558, A33

\bibitem[{{Baratta} \& {Strazzulla}(1990)}]{Baratta1990}
{Baratta}, G.~A. \& {Strazzulla}, G. 1990, \aap, 240, 429

\bibitem[{{Benjamin} {et~al.}(2003){Benjamin}, {Churchwell}, {Babler}, {Bania},
  {Clemens}, {Cohen}, {Dickey}, {Indebetouw}, {Jackson}, {Kobulnicky},
  {Lazarian}, {Marston}, {Mathis}, {Meade}, {Seager}, {Stolovy}, {Watson},
  {Whitney}, {Wolff}, \& {Wolfire}}]{Benjamin2003}
{Benjamin}, R.~A., {Churchwell}, E., {Babler}, B.~L., {et~al.} 2003, \pasp,
  115, 953

\bibitem[{{Bohlin} {et~al.}(2014){Bohlin}, {Gordon}, \&
  {Tremblay}}]{Bohlin2014}
{Bohlin}, R.~C., {Gordon}, K.~D., \& {Tremblay}, P.~E. 2014, \pasp, 126, 711

\bibitem[{{Bonnarel} {et~al.}(2000){Bonnarel}, {Fernique}, {Bienaym{\'e}},
  {Egret}, {Genova}, {Louys}, {Ochsenbein}, {Wenger}, \&
  {Bartlett}}]{Bonnarel2000}
{Bonnarel}, F., {Fernique}, P., {Bienaym{\'e}}, O., {et~al.} 2000, \aaps, 143,
  33

\bibitem[{{Boogert} {et~al.}(2013){Boogert}, {Chiar}, {Knez}, {{\"O}berg},
  {Mundy}, {Pendleton}, {Tielens}, \& {van Dishoeck}}]{Boogert2013}
{Boogert}, A.~C.~A., {Chiar}, J.~E., {Knez}, C., {et~al.} 2013, \apj, 777, 73

\bibitem[{{Boogert} {et~al.}(2015){Boogert}, {Gerakines}, \&
  {Whittet}}]{Boogert2015}
{Boogert}, A.~C.~A., {Gerakines}, P.~A., \& {Whittet}, D. C.~B. 2015, \araa,
  53, 541

\bibitem[{{Boogert} {et~al.}(2011){Boogert}, {Huard}, {Cook}, {Chiar}, {Knez},
  {Decin}, {Blake}, {Tielens}, \& {van Dishoeck}}]{Boogert2011}
{Boogert}, A.~C.~A., {Huard}, T.~L., {Cook}, A.~M., {et~al.} 2011, \apj, 729,
  92

\bibitem[{{Chiar} {et~al.}(2011){Chiar}, {Pendleton}, {Allamandola}, {Boogert},
  {Ennico}, {Greene}, {Geballe}, {Keane}, {Lada}, {Mason}, {Roellig},
  {Sandford}, {Tielens}, {Werner}, {Whittet}, {Decin}, \&
  {Eriksson}}]{Chiar2011}
{Chiar}, J.~E., {Pendleton}, Y.~J., {Allamandola}, L.~J., {et~al.} 2011, \apj,
  731, 9

\bibitem[{{Crill} {et~al.}(2020){Crill}, {Werner}, {Akeson}, {Ashby}, {Bleem},
  {Bock}, {Bryan}, {Burnham}, {Byunh}, {Chang}, {Chiang}, {Cook}, {Cooray},
  {Davis}, {Dor{\'e}}, {Dowell}, {Dubois-Felsmann}, {Eifler}, {Faisst},
  {Habib}, {Heinrich}, {Heitmann}, {Heaton}, {Hirata}, {Hristov}, {Hui},
  {Jeong}, {Kang}, {Kecman}, {Kirkpatrick}, {Korngut}, {Krause}, {Lee},
  {Lisse}, {Masters}, {Mauskopf}, {Melnick}, {Miyasaka}, {Nayyeri}, {Nguyen},
  {{\"O}berg}, {Padin}, {Paladini}, {Pourrahmani}, {Pyo}, {Smith}, {Song},
  {Symons}, {Teplitz}, {Tolls}, {Unwin}, {Windhorst}, {Yang}, \&
  {Zemcov}}]{Crill2020}
{Crill}, B.~P., {Werner}, M., {Akeson}, R., {et~al.} 2020, in Society of
  Photo-Optical Instrumentation Engineers (SPIE) Conference Series, Vol. 11443,
  Space Telescopes and Instrumentation 2020: Optical, Infrared, and Millimeter
  Wave, ed. M.~{Lystrup} \& M.~D. {Perrin}, 114430I

\bibitem[{{Cutri} {et~al.}(2021){Cutri}, {Wright}, {Conrow}, {Fowler},
  {Eisenhardt}, {Grillmair}, {Kirkpatrick}, {Masci}, {McCallon}, {Wheelock},
  {Fajardo-Acosta}, {Yan}, {Benford}, {Harbut}, {Jarrett}, {Lake}, {Leisawitz},
  {Ressler}, {Stanford}, {Tsai}, {Liu}, {Helou}, {Mainzer}, {Gettngs},
  {Gonzalez}, {Hoffman}, {Marsh}, {Padgett}, {Skrutskie}, {Beck}, {Papin}, \&
  {Wittman}}]{Cutri2014}
{Cutri}, R.~M., {Wright}, E.~L., {Conrow}, T., {et~al.} 2021, {VizieR Online
  Data Catalog: AllWISE Data Release (Cutri+ 2013)}, VizieR On-line Data
  Catalog: II/328. Originally published in: IPAC/Caltech (2013)

\bibitem[{{Danielson} {et~al.}(1965){Danielson}, {Woolf}, \&
  {Gaustad}}]{Danielson1965}
{Danielson}, R.~E., {Woolf}, N.~J., \& {Gaustad}, J.~E. 1965, \apj, 141, 116

\bibitem[{{Dartois} {et~al.}(2024){Dartois}, {Noble}, {Caselli}, {Fraser},
  {Jim{\'e}nez-Serra}, {Mat{\'e}}, {McClure}, {Melnick}, {Pendleton},
  {Shimonishi}, {Smith}, {Sturm}, {Taillard}, {Wakelam}, {Boogert},
  {Drozdovskaya}, {Erkal}, {Harsono}, {Herrero}, {Ioppolo}, {Linnartz},
  {McGuire}, {Perotti}, {Qasim}, \& {Rocha}}]{Dartois2024}
{Dartois}, E., {Noble}, J.~A., {Caselli}, P., {et~al.} 2024, Nature Astronomy,
  8, 359

\bibitem[{{Draine}(2003)}]{Draine2003}
{Draine}, B.~T. 2003, \araa, 41, 241

\bibitem[{{Evans} {et~al.}(2003){Evans}, {Allen}, {Blake}, {Boogert}, {Bourke},
  {Harvey}, {Kessler}, {Koerner}, {Lee}, {Mundy}, {Myers}, {Padgett},
  {Pontoppidan}, {Sargent}, {Stapelfeldt}, {van Dishoeck}, {Young}, \&
  {Young}}]{Evans2003}
{Evans}, II, N.~J., {Allen}, L.~E., {Blake}, G.~A., {et~al.} 2003, \pasp, 115,
  965

\bibitem[{{Fazio} {et~al.}(2004){Fazio}, {Hora}, {Allen}, {Ashby}, {Barmby},
  {Deutsch}, {Huang}, {Kleiner}, {Marengo}, {Megeath}, {Melnick}, {Pahre},
  {Patten}, {Polizotti}, {Smith}, {Taylor}, {Wang}, {Willner}, {Hoffmann},
  {Pipher}, {Forrest}, {McMurty}, {McCreight}, {McKelvey}, {McMurray}, {Koch},
  {Moseley}, {Arendt}, {Mentzell}, {Marx}, {Losch}, {Mayman}, {Eichhorn},
  {Krebs}, {Jhabvala}, {Gezari}, {Fixsen}, {Flores}, {Shakoorzadeh}, {Jungo},
  {Hakun}, {Workman}, {Karpati}, {Kichak}, {Whitley}, {Mann}, {Tollestrup},
  {Eisenhardt}, {Stern}, {Gorjian}, {Bhattacharya}, {Carey}, {Nelson},
  {Glaccum}, {Lacy}, {Lowrance}, {Laine}, {Reach}, {Stauffer}, {Surace},
  {Wilson}, {Wright}, {Hoffman}, {Domingo}, \& {Cohen}}]{Fazio2004}
{Fazio}, G.~G., {Hora}, J.~L., {Allen}, L.~E., {et~al.} 2004, \apjs, 154, 10

\bibitem[{Fouesneau(2025)}]{pyphot}
Fouesneau, M. 2025, pyphot, Zenodo, Version pyphot\_v1.6.0

\bibitem[{{Gaia Collaboration} {et~al.}(2016){Gaia Collaboration}, {Prusti},
  {de Bruijne}, {Brown}, {Vallenari}, {Babusiaux}, {Bailer-Jones}, {Bastian},
  {Biermann}, {Evans}, {Eyer}, {Jansen}, {Jordi}, {Klioner}, {Lammers},
  {Lindegren}, {Luri}, {Mignard}, {Milligan}, {Panem}, {Poinsignon},
  {Pourbaix}, {Randich}, {Sarri}, {Sartoretti}, {Siddiqui}, {Soubiran},
  {Valette}, {van Leeuwen}, {Walton}, {Aerts}, {Arenou}, {Cropper}, {Drimmel},
  {H{\o}g}, {Katz}, {Lattanzi}, {O'Mullane}, {Grebel}, {Holland}, {Huc},
  {Passot}, {Bramante}, {Cacciari}, {Casta{\~n}eda}, {Chaoul}, {Cheek}, {De
  Angeli}, {Fabricius}, {Guerra}, {Hern{\'a}ndez}, {Jean-Antoine-Piccolo},
  {Masana}, {Messineo}, {Mowlavi}, {Nienartowicz}, {Ord{\'o}{\~n}ez-Blanco},
  {Panuzzo}, {Portell}, {Richards}, {Riello}, {Seabroke}, {Tanga},
  {Th{\'e}venin}, {Torra}, {Els}, {Gracia-Abril}, {Comoretto},
  {Garcia-Reinaldos}, {Lock}, {Mercier}, {Altmann}, {Andrae}, {Astraatmadja},
  {Bellas-Velidis}, {Benson}, {Berthier}, {Blomme}, {Busso}, {Carry},
  {Cellino}, {Clementini}, {Cowell}, {Creevey}, {Cuypers}, {Davidson}, {De
  Ridder}, {de Torres}, {Delchambre}, {Dell'Oro}, {Ducourant}, {Fr{\'e}mat},
  {Garc{\'\i}a-Torres}, {Gosset}, {Halbwachs}, {Hambly}, {Harrison}, {Hauser},
  {Hestroffer}, {Hodgkin}, {Huckle}, {Hutton}, {Jasniewicz}, {Jordan},
  {Kontizas}, {Korn}, {Lanzafame}, {Manteiga}, {Moitinho}, {Muinonen},
  {Osinde}, {Pancino}, {Pauwels}, {Petit}, {Recio-Blanco}, {Robin}, {Sarro},
  {Siopis}, {Smith}, {Smith}, {Sozzetti}, {Thuillot}, {van Reeven}, {Viala},
  {Abbas}, {Abreu Aramburu}, {Accart}, {Aguado}, {Allan}, {Allasia},
  {Altavilla}, {{\'A}lvarez}, {Alves}, {Anderson}, {Andrei}, {Anglada Varela},
  {Antiche}, {Antoja}, {Ant{\'o}n}, {Arcay}, {Atzei}, {Ayache}, {Bach},
  {Baker}, {Balaguer-N{\'u}{\~n}ez}, {Barache}, {Barata}, {Barbier}, {Barblan},
  {Baroni}, {Barrado y Navascu{\'e}s}, {Barros}, {Barstow}, {Becciani},
  {Bellazzini}, {Bellei}, {Bello Garc{\'\i}a}, {Belokurov}, {Bendjoya},
  {Berihuete}, {Bianchi}, {Bienaym{\'e}}, {Billebaud}, {Blagorodnova},
  {Blanco-Cuaresma}, {Boch}, {Bombrun}, {Borrachero}, {Bouquillon}, {Bourda},
  {Bouy}, {Bragaglia}, {Breddels}, {Brouillet}, {Br{\"u}semeister},
  {Bucciarelli}, {Budnik}, {Burgess}, {Burgon}, {Burlacu}, {Busonero}, {Buzzi},
  {Caffau}, {Cambras}, {Campbell}, {Cancelliere}, {Cantat-Gaudin}, {Carlucci},
  {Carrasco}, {Castellani}, {Charlot}, {Charnas}, {Charvet}, {Chassat},
  {Chiavassa}, {Clotet}, {Cocozza}, {Collins}, {Collins}, \& {Costigan}}]{Gaia}
{Gaia Collaboration}, {Prusti}, T., {de Bruijne}, J.~H.~J., {et~al.} 2016,
  \aap, 595, A1

\bibitem[{{Gaia Collaboration} {et~al.}(2023){Gaia Collaboration}, {Vallenari},
  {Brown}, {Prusti}, {de Bruijne}, {Arenou}, {Babusiaux}, {Biermann},
  {Creevey}, {Ducourant}, {Evans}, {Eyer}, {Guerra}, {Hutton}, {Jordi},
  {Klioner}, {Lammers}, {Lindegren}, {Luri}, {Mignard}, {Panem}, {Pourbaix},
  {Randich}, {Sartoretti}, {Soubiran}, {Tanga}, {Walton}, {Bailer-Jones},
  {Bastian}, {Drimmel}, {Jansen}, {Katz}, {Lattanzi}, {van Leeuwen}, {Bakker},
  {Cacciari}, {Casta{\~n}eda}, {De Angeli}, {Fabricius}, {Fouesneau},
  {Fr{\'e}mat}, {Galluccio}, {Guerrier}, {Heiter}, {Masana}, {Messineo},
  {Mowlavi}, {Nicolas}, {Nienartowicz}, {Pailler}, {Panuzzo}, {Riclet}, {Roux},
  {Seabroke}, {Sordo}, {Th{\'e}venin}, {Gracia-Abril}, {Portell}, {Teyssier},
  {Altmann}, {Andrae}, {Audard}, {Bellas-Velidis}, {Benson}, {Berthier},
  {Blomme}, {Burgess}, {Busonero}, {Busso}, {C{\'a}novas}, {Carry}, {Cellino},
  {Cheek}, {Clementini}, {Damerdji}, {Davidson}, {de Teodoro}, {Nu{\~n}ez
  Campos}, {Delchambre}, {Dell'Oro}, {Esquej}, {Fern{\'a}ndez-Hern{\'a}ndez},
  {Fraile}, {Garabato}, {Garc{\'\i}a-Lario}, {Gosset}, {Haigron}, {Halbwachs},
  {Hambly}, {Harrison}, {Hern{\'a}ndez}, {Hestroffer}, {Hodgkin}, {Holl},
  {Jan{\ss}en}, {Jevardat de Fombelle}, {Jordan}, {Krone-Martins}, {Lanzafame},
  {L{\"o}ffler}, {Marchal}, {Marrese}, {Moitinho}, {Muinonen}, {Osborne},
  {Pancino}, {Pauwels}, {Recio-Blanco}, {Reyl{\'e}}, {Riello}, {Rimoldini},
  {Roegiers}, {Rybizki}, {Sarro}, {Siopis}, {Smith}, {Sozzetti}, {Utrilla},
  {van Leeuwen}, {Abbas}, {{\'A}brah{\'a}m}, {Abreu Aramburu}, {Aerts},
  {Aguado}, {Ajaj}, {Aldea-Montero}, {Altavilla}, {{\'A}lvarez}, {Alves},
  {Anders}, {Anderson}, {Anglada Varela}, {Antoja}, {Baines}, {Baker},
  {Balaguer-N{\'u}{\~n}ez}, {Balbinot}, {Balog}, {Barache}, {Barbato},
  {Barros}, {Barstow}, {Bartolom{\'e}}, {Bassilana}, {Bauchet}, {Becciani},
  {Bellazzini}, {Berihuete}, {Bernet}, {Bertone}, {Bianchi}, {Binnenfeld},
  {Blanco-Cuaresma}, {Blazere}, {Boch}, {Bombrun}, {Bossini}, {Bouquillon},
  {Bragaglia}, {Bramante}, {Breedt}, {Bressan}, {Brouillet}, {Brugaletta},
  {Bucciarelli}, {Burlacu}, {Butkevich}, {Buzzi}, {Caffau}, {Cancelliere},
  {Cantat-Gaudin}, {Carballo}, {Carlucci}, {Carnerero}, {Carrasco},
  {Casamiquela}, {Castellani}, {Castro-Ginard}, {Chaoul}, {Charlot}, {Chemin},
  {Chiaramida}, {Chiavassa}, {Chornay}, {Comoretto}, {Contursi}, {Cooper},
  {Cornez}, {Cowell}, {Crifo}, {Cropper}, {Crosta}, {Crowley}, {Dafonte},
  {Dapergolas}, {David}, {David}, {de Laverny}, {De Luise}, \& {De
  March}}]{GaiaDR3}
{Gaia Collaboration}, {Vallenari}, A., {Brown}, A.~G.~A., {et~al.} 2023, \aap,
  674, A1

\bibitem[{{Gardner} {et~al.}(2006){Gardner}, {Mather}, {Clampin}, {Doyon},
  {Greenhouse}, {Hammel}, {Hutchings}, {Jakobsen}, {Lilly}, {Long}, {Lunine},
  {McCaughrean}, {Mountain}, {Nella}, {Rieke}, {Rieke}, {Rix}, {Smith},
  {Sonneborn}, {Stiavelli}, {Stockman}, {Windhorst}, \& {Wright}}]{Gardner2006}
{Gardner}, J.~P., {Mather}, J.~C., {Clampin}, M., {et~al.} 2006, \ssr, 123, 485

\bibitem[{{Gillett} \& {Forrest}(1973)}]{Gillet1973}
{Gillett}, F.~C. \& {Forrest}, W.~J. 1973, \apj, 179, 483

\bibitem[{{Ginsburg} {et~al.}(2019){Ginsburg}, {Sip{\H{o}}cz}, {Brasseur},
  {Cowperthwaite}, {Craig}, {Deil}, {Guillochon}, {Guzman}, {Liedtke}, {Lian
  Lim}, {Lockhart}, {Mommert}, {Morris}, {Norman}, {Parikh}, {Persson},
  {Robitaille}, {Segovia}, {Singer}, {Tollerud}, {de Val-Borro}, {Valtchanov},
  {Woillez}, {Astroquery Collaboration}, \& {a subset of astropy
  Collaboration}}]{astroquery}
{Ginsburg}, A., {Sip{\H{o}}cz}, B.~M., {Brasseur}, C.~E., {et~al.} 2019, \aj,
  157, 98

\bibitem[{{Goodman} {et~al.}(2009){Goodman}, {Pineda}, \&
  {Schnee}}]{Goodman2009}
{Goodman}, A.~A., {Pineda}, J.~E., \& {Schnee}, S.~L. 2009, \apj, 692, 91

\bibitem[{{Goto} {et~al.}(2018){Goto}, {Bailey}, {Hocuk}, {Caselli},
  {Esplugues}, {Cazaux}, \& {Spaans}}]{Goto2018}
{Goto}, M., {Bailey}, J.~D., {Hocuk}, S., {et~al.} 2018, \aap, 610, A9

\bibitem[{{Green} {et~al.}(2024){Green}, {Zhang}, \& {Zhang}}]{Green2024}
{Green}, G.~M., {Zhang}, X., \& {Zhang}, R. 2024, arXiv e-prints,
  arXiv:2410.22537

\bibitem[{{Greene} {et~al.}(2017){Greene}, {Kelly}, {Stansberry}, {Leisenring},
  {Egami}, {Schlawin}, {Chu}, {Hodapp}, \& {Rieke}}]{Greene2017}
{Greene}, T.~P., {Kelly}, D.~M., {Stansberry}, J., {et~al.} 2017, Journal of
  Astronomical Telescopes, Instruments, and Systems, 3, 035001

\bibitem[{Harris {et~al.}(2020)Harris, Millman, van~der Walt, Gommers,
  Virtanen, Cournapeau, Wieser, Taylor, Berg, Smith, Kern, Picus, Hoyer, van
  Kerkwijk, Brett, Haldane, del R{\'{i}}o, Wiebe, Peterson,
  G{\'{e}}rard-Marchant, Sheppard, Reddy, Weckesser, Abbasi, Gohlke, \&
  Oliphant}]{numpy}
Harris, C.~R., Millman, K.~J., van~der Walt, S.~J., {et~al.} 2020, Nature, 585,
  357

\bibitem[{{Hasenberger} {et~al.}(2018){Hasenberger}, {Lombardi}, {Alves},
  {Forbrich}, {Hacar}, \& {Lada}}]{Hasenberger2018}
{Hasenberger}, B., {Lombardi}, M., {Alves}, J., {et~al.} 2018, \aap, 620, A24

\bibitem[{{Houck} {et~al.}(2004){Houck}, {Roellig}, {van Cleve}, {Forrest},
  {Herter}, {Lawrence}, {Matthews}, {Reitsema}, {Soifer}, {Watson}, {Weedman},
  {Huisjen}, {Troeltzsch}, {Barry}, {Bernard-Salas}, {Blacken}, {Brandl},
  {Charmandaris}, {Devost}, {Gull}, {Hall}, {Henderson}, {Higdon}, {Pirger},
  {Schoenwald}, {Sloan}, {Uchida}, {Appleton}, {Armus}, {Burgdorf},
  {Fajardo-Acosta}, {Grillmair}, {Ingalls}, {Morris}, \& {Teplitz}}]{Houck2004}
{Houck}, J.~R., {Roellig}, T.~L., {van Cleve}, J., {et~al.} 2004, \apjs, 154,
  18

\bibitem[{{Hudgins} {et~al.}(1993){Hudgins}, {Sandford}, {Allamandola}, \&
  {Tielens}}]{Hudgins1993}
{Hudgins}, D.~M., {Sandford}, S.~A., {Allamandola}, L.~J., \& {Tielens},
  A.~G.~G.~M. 1993, \apjs, 86, 713

\bibitem[{Hunter(2007)}]{matplotlib}
Hunter, J.~D. 2007, Computing in Science \& Engineering, 9, 90

\bibitem[{{Jakobsen} {et~al.}(2022){Jakobsen}, {Ferruit}, {Alves de Oliveira},
  {Arribas}, {Bagnasco}, {Barho}, {Beck}, {Birkmann}, {B{\"o}ker}, {Bunker},
  {Charlot}, {de Jong}, {de Marchi}, {Ehrenwinkler}, {Falcolini}, {Fels},
  {Franx}, {Franz}, {Funke}, {Giardino}, {Gnata}, {Holota}, {Honnen}, {Jensen},
  {Jentsch}, {Johnson}, {Jollet}, {Karl}, {Kling}, {K{\"o}hler}, {Kolm},
  {Kumari}, {Lander}, {Lemke}, {L{\'o}pez-Caniego}, {L{\"u}tzgendorf},
  {Maiolino}, {Manjavacas}, {Marston}, {Maschmann}, {Maurer}, {Messerschmidt},
  {Moseley}, {Mosner}, {Mott}, {Muzerolle}, {Pirzkal}, {Pittet}, {Plitzke},
  {Posselt}, {Rapp}, {Rauscher}, {Rawle}, {Rix}, {R{\"o}del}, {Rumler},
  {Sabbi}, {Salvignol}, {Schmid}, {Sirianni}, {Smith}, {Strada}, {te Plate},
  {Valenti}, {Wettemann}, {Wiehe}, {Wiesmayer}, {Willott}, {Wright}, {Zeidler},
  \& {Zincke}}]{Jakobsen2022}
{Jakobsen}, P., {Ferruit}, P., {Alves de Oliveira}, C., {et~al.} 2022, \aap,
  661, A80

\bibitem[{{Johnson} \& {Morgan}(1953)}]{Johnson1953}
{Johnson}, H.~L. \& {Morgan}, W.~W. 1953, \apj, 117, 313

\bibitem[{{Knacke} {et~al.}(1969){Knacke}, {Cudaback}, \&
  {Gaustad}}]{Knacke1969}
{Knacke}, R.~F., {Cudaback}, D.~D., \& {Gaustad}, J.~E. 1969, \apj, 158, 151

\bibitem[{{Lada} {et~al.}(1994){Lada}, {Lada}, {Clemens}, \&
  {Bally}}]{Lada1994}
{Lada}, C.~J., {Lada}, E.~A., {Clemens}, D.~P., \& {Bally}, J. 1994, \apj, 429,
  694

\bibitem[{{Lada} {et~al.}(2017){Lada}, {Lewis}, {Lombardi}, \&
  {Alves}}]{Lada2017}
{Lada}, C.~J., {Lewis}, J.~A., {Lombardi}, M., \& {Alves}, J. 2017, \aap, 606,
  A100

\bibitem[{{Lombardi}(2018)}]{Lombardi2018}
{Lombardi}, M. 2018, \aap, 615, A174

\bibitem[{{Lombardi} \& {Alves}(2001)}]{Lombardi2001}
{Lombardi}, M. \& {Alves}, J. 2001, \aap, 377, 1023

\bibitem[{{Lombardi} {et~al.}(2014){Lombardi}, {Bouy}, {Alves}, \&
  {Lada}}]{Lombardi2014}
{Lombardi}, M., {Bouy}, H., {Alves}, J., \& {Lada}, C.~J. 2014, \aap, 566, A45

\bibitem[{{Madden} {et~al.}(2022){Madden}, {Boogert}, {Chiar}, {Knez},
  {Pendleton}, {Tielens}, \& {Yip}}]{Madden2022}
{Madden}, M.~C.~L., {Boogert}, A.~C.~A., {Chiar}, J.~E., {et~al.} 2022, \apj,
  930, 2

\bibitem[{{Mainzer} {et~al.}(2011){Mainzer}, {Bauer}, {Grav}, {Masiero},
  {Cutri}, {Dailey}, {Eisenhardt}, {McMillan}, {Wright}, {Walker}, {Jedicke},
  {Spahr}, {Tholen}, {Alles}, {Beck}, {Brandenburg}, {Conrow}, {Evans},
  {Fowler}, {Jarrett}, {Marsh}, {Masci}, {McCallon}, {Wheelock}, {Wittman},
  {Wyatt}, {DeBaun}, {Elliott}, {Elsbury}, {Gautier}, {Gomillion}, {Leisawitz},
  {Maleszewski}, {Micheli}, \& {Wilkins}}]{Mainzer2011}
{Mainzer}, A., {Bauer}, J., {Grav}, T., {et~al.} 2011, \apj, 731, 53

\bibitem[{{Majewski} {et~al.}(2011){Majewski}, {Zasowski}, \&
  {Nidever}}]{Majewski2011}
{Majewski}, S.~R., {Zasowski}, G., \& {Nidever}, D.~L. 2011, \apj, 739, 25

\bibitem[{{McClure} {et~al.}(2023){McClure}, {Rocha}, {Pontoppidan}, {Crouzet},
  {Chu}, {Dartois}, {Lamberts}, {Noble}, {Pendleton}, {Perotti}, {Qasim},
  {Rachid}, {Smith}, {Sun}, {Beck}, {Boogert}, {Brown}, {Caselli}, {Charnley},
  {Cuppen}, {Dickinson}, {Drozdovskaya}, {Egami}, {Erkal}, {Fraser}, {Garrod},
  {Harsono}, {Ioppolo}, {Jim{\'e}nez-Serra}, {Jin}, {J{\o}rgensen},
  {Kristensen}, {Lis}, {McCoustra}, {McGuire}, {Melnick}, {{\~A}-berg},
  {Palumbo}, {Shimonishi}, {Sturm}, {van Dishoeck}, \&
  {Linnartz}}]{McClure2023}
{McClure}, M.~K., {Rocha}, W.~R.~M., {Pontoppidan}, K.~M., {et~al.} 2023,
  Nature Astronomy, 7, 431

\bibitem[{McKinney(2010)}]{Pandas2}
McKinney, W. 2010, in Proceedings of the 9th Python in Science Conference, ed.
  S.~van~der Walt \& J.~Millman, 56--61

\bibitem[{{Meingast} {et~al.}(2023){Meingast}, {Alves}, {Bouy}, {Petr-Gotzens},
  {F{\"u}rnkranz}, {Gro{\ss}schedl}, {Hernandez}, {Rottensteiner}, {Arnaboldi},
  {Ascenso}, {Bayo}, {Br{\"a}ndli}, {Brown}, {Forbrich}, {Goodman}, {Hacar},
  {Hasenberger}, {K{\"o}hler}, {Kubiak}, {Kuhn}, {Lada}, {Leschinski},
  {Lombardi}, {Mardones}, {Mascetti}, {Miret-Roig}, {Moitinho},
  {Mu{\v{z}}i{\'c}}, {Piecka}, {Posch}, {Prusti}, {Pe{\~n}a Ram{\'\i}rez},
  {Ramlau}, {Ratzenb{\"o}ck}, {Sacco}, {Swiggum}, {Teixeira}, {Urban}, {Zari},
  \& {Zucker}}]{Meingast2023}
{Meingast}, S., {Alves}, J., {Bouy}, H., {et~al.} 2023, \aap, 673, A58

\bibitem[{{Meingast} {et~al.}(2018){Meingast}, {Alves}, \&
  {Lombardi}}]{Meingast2018}
{Meingast}, S., {Alves}, J., \& {Lombardi}, M. 2018, \aap, 614, A65

\bibitem[{{Meingast} {et~al.}(2017){Meingast}, {Lombardi}, \&
  {Alves}}]{Meingast2017}
{Meingast}, S., {Lombardi}, M., \& {Alves}, J. 2017, \aap, 601, A137

\bibitem[{{Meisner} {et~al.}(2023){Meisner}, {Caselden}, {Schlafly}, \&
  {Kiwy}}]{Meisner2023}
{Meisner}, A.~M., {Caselden}, D., {Schlafly}, E.~F., \& {Kiwy}, F. 2023, \aj,
  165, 36

\bibitem[{{Minniti} {et~al.}(2010){Minniti}, {Lucas}, {Emerson}, {Saito},
  {Hempel}, {Pietrukowicz}, {Ahumada}, {Alonso}, {Alonso-Garcia}, {Arias},
  {Bandyopadhyay}, {Barb{\'a}}, {Barbuy}, {Bedin}, {Bica}, {Borissova},
  {Bronfman}, {Carraro}, {Catelan}, {Clari{\'a}}, {Cross}, {de Grijs},
  {D{\'e}k{\'a}ny}, {Drew}, {Fari{\~n}a}, {Feinstein}, {Fern{\'a}ndez
  Laj{\'u}s}, {Gamen}, {Geisler}, {Gieren}, {Goldman}, {Gonzalez}, {Gunthardt},
  {Gurovich}, {Hambly}, {Irwin}, {Ivanov}, {Jord{\'a}n}, {Kerins}, {Kinemuchi},
  {Kurtev}, {L{\'o}pez-Corredoira}, {Maccarone}, {Masetti}, {Merlo},
  {Messineo}, {Mirabel}, {Monaco}, {Morelli}, {Padilla}, {Palma}, {Parisi},
  {Pignata}, {Rejkuba}, {Roman-Lopes}, {Sale}, {Schreiber}, {Schr{\"o}der},
  {Smith}, {}, {Soto}, {Tamura}, {Tappert}, {Thompson}, {Toledo}, {Zoccali}, \&
  {Pietrzynski}}]{Minniti2010}
{Minniti}, D., {Lucas}, P.~W., {Emerson}, J.~P., {et~al.} 2010, \na, 15, 433

\bibitem[{{Murakawa} {et~al.}(2000){Murakawa}, {Tamura}, \&
  {Nagata}}]{Murakawa2000}
{Murakawa}, K., {Tamura}, M., \& {Nagata}, T. 2000, \apjs, 128, 603

\bibitem[{{Noble} {et~al.}(2013){Noble}, {Fraser}, {Aikawa}, {Pontoppidan}, \&
  {Sakon}}]{Noble2013}
{Noble}, J.~A., {Fraser}, H.~J., {Aikawa}, Y., {Pontoppidan}, K.~M., \&
  {Sakon}, I. 2013, \apj, 775, 85

\bibitem[{{Noble} {et~al.}(2024){Noble}, {Fraser}, {Smith}, {Dartois},
  {Boogert}, {Cuppen}, {Dickinson}, {Dulieu}, {Egami}, {Erkal}, {Giuliano},
  {Husquinet}, {Lamberts}, {Mat{\'e}}, {McClure}, {Palumbo}, {Shimonishi},
  {Sun}, {Bergner}, {Brown}, {Caselli}, {Congiu}, {Drozdovskaya}, {Herrero},
  {Ioppolo}, {Jimenez-Serra}, {Linnartz}, {Melnick}, {McGuire}, {Oberg},
  {Perotti}, {Qasim}, {Rocha}, \& {Urso}}]{Noble2024}
{Noble}, J.~A., {Fraser}, H.~J., {Smith}, Z.~L., {et~al.} 2024, Nature
  Astronomy, 8, 1169

\bibitem[{Ochsenbein(1996)}]{10.26093/cds/vizier}
Ochsenbein, F. 1996, The VizieR database of astronomical catalogues

\bibitem[{{Ochsenbein} {et~al.}(2000){Ochsenbein}, {Bauer}, \&
  {Marcout}}]{vizier2000}
{Ochsenbein}, F., {Bauer}, P., \& {Marcout}, J. 2000, \aaps, 143, 23

\bibitem[{{Ohyama} {et~al.}(2007){Ohyama}, {Onaka}, {Matsuhara}, {Wada}, {Kim},
  {Fujishiro}, {Uemizu}, {Sakon}, {Cohen}, {Ishigaki}, {Ishihara}, {Ita},
  {Kataza}, {Matsumoto}, {Murakami}, {Oyabu}, {Tanab{\'e}}, {Takagi}, {Ueno},
  {Usui}, {Watarai}, {Pearson}, {Takeyama}, {Yamamuro}, \&
  {Ikeda}}]{Ohyama2007}
{Ohyama}, Y., {Onaka}, T., {Matsuhara}, H., {et~al.} 2007, \pasj, 59, S411

\bibitem[{{Onaka} {et~al.}(2007){Onaka}, {Matsuhara}, {Wada}, {Fujishiro},
  {Fujiwara}, {Ishigaki}, {Ishihara}, {Ita}, {Kataza}, {Kim}, {Matsumoto},
  {Murakami}, {Ohyama}, {Oyabu}, {Sakon}, {Tanab{\'e}}, {Takagi}, {Uemizu},
  {Ueno}, {Usui}, {Watarai}, {Cohen}, {Enya}, {Ootsubo}, {Pearson}, {Takeyama},
  {Yamamuro}, \& {Ikeda}}]{Onaka2007}
{Onaka}, T., {Matsuhara}, H., {Wada}, T., {et~al.} 2007, \pasj, 59, S401

\bibitem[{{Pandas Development Team}(2020)}]{Pandas1}
{Pandas Development Team}. 2020, pandas-dev/pandas: Pandas

\bibitem[{Pedregosa {et~al.}(2011)Pedregosa, Varoquaux, Gramfort, Michel,
  Thirion, Grisel, Blondel, Prettenhofer, Weiss, Dubourg, Vanderplas, Passos,
  Cournapeau, Brucher, Perrot, \& {{\'E}}douard Duchesnay}]{scikit-learn}
Pedregosa, F., Varoquaux, G., Gramfort, A., {et~al.} 2011, Journal of Machine
  Learning Research, 12, 2825

\bibitem[{{Persson} {et~al.}(1998){Persson}, {Murphy}, {Krzeminski}, {Roth}, \&
  {Rieke}}]{Persson1998}
{Persson}, S.~E., {Murphy}, D.~C., {Krzeminski}, W., {Roth}, M., \& {Rieke},
  M.~J. 1998, \aj, 116, 2475

\bibitem[{{Pilbratt} {et~al.}(2010){Pilbratt}, {Riedinger}, {Passvogel},
  {Crone}, {Doyle}, {Gageur}, {Heras}, {Jewell}, {Metcalfe}, {Ott}, \&
  {Schmidt}}]{Herschel}
{Pilbratt}, G.~L., {Riedinger}, J.~R., {Passvogel}, T., {et~al.} 2010, \aap,
  518, L1

\bibitem[{{Planck Collaboration} {et~al.}(2011){Planck Collaboration}, {Ade},
  {Aghanim}, {Arnaud}, {Ashdown}, {Aumont}, {Baccigalupi}, {Baker}, {Balbi},
  {Banday}, {Barreiro}, {Bartlett}, {Battaner}, {Benabed}, {Bennett},
  {Beno{\^\i}t}, {Bernard}, {Bersanelli}, {Bhatia}, {Bock}, {Bonaldi}, {Bond},
  {Borrill}, {Bouchet}, {Bradshaw}, {Bremer}, {Bucher}, {Burigana}, {Butler},
  {Cabella}, {Cantalupo}, {Cappellini}, {Cardoso}, {Carr}, {Casale},
  {Catalano}, {Cay{\'o}n}, {Challinor}, {Chamballu}, {Charra}, {Chary},
  {Chiang}, {Chiang}, {Christensen}, {Clements}, {Colombi}, {Couchot},
  {Coulais}, {Crill}, {Crone}, {Crook}, {Cuttaia}, {Danese}, {D'Arcangelo},
  {Davies}, {Davis}, {de Bernardis}, {de Bruin}, {de Gasperis}, {de Rosa}, {de
  Zotti}, {Delabrouille}, {Delouis}, {D{\'e}sert}, {Dick}, {Dickinson},
  {Dolag}, {Dole}, {Donzelli}, {Dor{\'e}}, {D{\"o}rl}, {Douspis}, {Dupac},
  {Efstathiou}, {En{\ss}lin}, {Eriksen}, {Finelli}, {Foley}, {Forni},
  {Fosalba}, {Frailis}, {Franceschi}, {Freschi}, {Gaier}, {Galeotta},
  {Gallegos}, {Gandolfo}, {Ganga}, {Giard}, {Giardino}, {Gienger},
  {Giraud-H{\'e}raud}, {Gonz{\'a}lez}, {Gonz{\'a}lez-Nuevo}, {G{\'o}rski},
  {Gratton}, {Gregorio}, {Gruppuso}, {Guyot}, {Haissinski}, {Hansen},
  {Harrison}, {Helou}, {Henrot-Versill{\'e}}, {Hern{\'a}ndez-Monteagudo},
  {Herranz}, {Hildebrandt}, {Hivon}, {Hobson}, {Holmes}, {Hornstrup}, {Hovest},
  {Hoyland}, {Huffenberger}, {Jaffe}, {Jagemann}, {Jones}, {Juillet}, {Juvela},
  {Kangaslahti}, {Keih{\"a}nen}, {Keskitalo}, {Kisner}, {Kneissl}, {Knox},
  {Krassenburg}, {Kurki-Suonio}, {Lagache}, {L{\"a}hteenm{\"a}ki}, {Lamarre},
  {Lange}, {Lasenby}, {Laureijs}, {Lawrence}, {Leach}, {Leahy}, {Leonardi},
  {Leroy}, {Lilje}, {Linden-V{\o}rnle}, {L{\'o}pez-Caniego}, {Lowe}, {Lubin},
  {Mac{\'\i}as-P{\'e}rez}, {Maciaszek}, {MacTavish}, {Maffei}, {Maino},
  {Mandolesi}, {Mann}, {Maris}, {Mart{\'\i}nez-Gonz{\'a}lez}, {Masi},
  {Massardi}, {Matarrese}, {Matthai}, {Mazzotta}, {McDonald}, {McGehee},
  {Meinhold}, {Melchiorri}, {Melin}, {Mendes}, {Mennella}, {Mevi},
  {Miniscalco}, {Mitra}, {Miville-Desch{\^e}nes}, {Moneti}, {Montier},
  {Morgante}, {Morisset}, {Mortlock}, {Munshi}, {Murphy}, {Naselsky}, {Natoli},
  {Netterfield}, {N{\o}rgaard-Nielsen}, {Noviello}, {Novikov}, {Novikov},
  {O'Dwyer}, {Ortiz}, {Osborne}, {Osuna}, {Oxborrow}, {Pajot}, {Paladini},
  {Partridge}, {Pasian}, {Passvogel}, {Patanchon}, {Pearson}, {Pearson},
  {Perdereau}, {Perotto}, {Perrotta}, {Piacentini}, {Piat}, {Pierpaoli},
  {Plaszczynski}, {Platania}, {Pointecouteau}, {Polenta}, {Ponthieu}, {Popa},
  {Poutanen}, {Pr{\'e}zeau}, {Prunet}, {Puget}, {Rachen}, {Reach}, {Rebolo},
  {Reinecke}, {Reix}, {Renault}, {Ricciardi}, {Riller}, {Ristorcelli}, {Rocha},
  {Rosset}, {Rowan-Robinson}, {Rubi{\~n}o-Mart{\'\i}n}, {Rusholme}, {Salerno},
  {Sandri}, {Santos}, {Savini}, {Schaefer}, {Scott}, {Seiffert}, {Shellard},
  {Simonetto}, {Smoot}, {Sozzi}, {Starck}, {Sternberg}, {Stivoli}, {Stolyarov},
  {Stompor}, {Stringhetti}, {Sudiwala}, {Sunyaev}, {Sygnet}, {Tapiador},
  {Tauber}, {Tavagnacco}, {Taylor}, {Terenzi}, {Texier}, {Toffolatti},
  {Tomasi}, {Torre}, {Tristram}, {Tuovinen}, {T{\"u}rler}, {Tuttlebee},
  {Umana}, {Valenziano}, {Valiviita}, {Varis}, {Vibert}, {Vielva}, {Villa},
  {Vittorio}, {Wade}, {Wandelt}, {Watson}, {White}, {White}, {Wilkinson},
  {Yvon}, {Zacchei}, \& {Zonca}}]{PlanckI}
{Planck Collaboration}, {Ade}, P.~A.~R., {Aghanim}, N., {et~al.} 2011, \aap,
  536, A1

\bibitem[{{Pokhrel} {et~al.}(2020){Pokhrel}, {Gutermuth}, {Betti}, {Offner},
  {Myers}, {Megeath}, {Sokol}, {Ali}, {Allen}, {Allen}, {Dunham}, {Fischer},
  {Henning}, {Heyer}, {Hora}, {Pipher}, {Tobin}, \& {Wolk}}]{Pokhrel2020}
{Pokhrel}, R., {Gutermuth}, R.~A., {Betti}, S.~K., {et~al.} 2020, \apj, 896, 60

\bibitem[{{Posch} {et~al.}(2023){Posch}, {Miret-Roig}, {Alves},
  {Ratzenb{\"o}ck}, {Gro{\ss}schedl}, {Meingast}, {Zucker}, \&
  {Burkert}}]{Posch2023}
{Posch}, L., {Miret-Roig}, N., {Alves}, J., {et~al.} 2023, \aap, 679, L10

\bibitem[{{Ratzenb{\"o}ck} {et~al.}(2023){Ratzenb{\"o}ck}, {Gro{\ss}schedl},
  {Alves}, {Miret-Roig}, {Bomze}, {Forbes}, {Goodman}, {Hacar}, {Lin},
  {Meingast}, {M{\"o}ller}, {Piecka}, {Posch}, {Rottensteiner}, {Swiggum}, \&
  {Zucker}}]{Ratzenboebck2023}
{Ratzenb{\"o}ck}, S., {Gro{\ss}schedl}, J.~E., {Alves}, J., {et~al.} 2023,
  \aap, 678, A71

\bibitem[{{Rayner} {et~al.}(2003){Rayner}, {Toomey}, {Onaka}, {Denault},
  {Stahlberger}, {Vacca}, {Cushing}, \& {Wang}}]{Rayner2003}
{Rayner}, J.~T., {Toomey}, D.~W., {Onaka}, P.~M., {et~al.} 2003, \pasp, 115,
  362

\bibitem[{{Rebull} {et~al.}(2010){Rebull}, {Padgett}, {McCabe}, {Hillenbrand},
  {Stapelfeldt}, {Noriega-Crespo}, {Carey}, {Brooke}, {Huard}, {Terebey},
  {Audard}, {Monin}, {Fukagawa}, {G{\"u}del}, {Knapp}, {Menard}, {Allen},
  {Angione}, {Baldovin-Saavedra}, {Bouvier}, {Briggs}, {Dougados}, {Evans},
  {Flagey}, {Guieu}, {Grosso}, {Glauser}, {Harvey}, {Hines}, {Latter},
  {Skinner}, {Strom}, {Tromp}, \& {Wolf}}]{Rebull2010}
{Rebull}, L.~M., {Padgett}, D.~L., {McCabe}, C.~E., {et~al.} 2010, \apjs, 186,
  259

\bibitem[{{Rocha} {et~al.}(2025){Rocha}, {McClure}, {Sturm}, {Beck}, {Smith},
  {Dickinson}, {Sun}, {Egami}, {Boogert}, {Fraser}, {Dartois}, {Jimenez-Serra},
  {Noble}, {Bergner}, {Caselli}, {Charnley}, {Chiar}, {Chu}, {Cooke},
  {Crouzet}, {van Dishoeck}, {Drozdovskaya}, {Garrod}, {Harsono}, {Ioppolo},
  {Jin}, {J{\o}rgensen}, {Lamberts}, {Lis}, {Melnick}, {McGuire}, {{\"O}berg},
  {Palumbo}, {Pendleton}, {Perotti}, {Qasim}, {Shope}, {Urso}, {Viti}, \&
  {Linnartz}}]{Rocha2025}
{Rocha}, W.~R.~M., {McClure}, M.~K., {Sturm}, J.~A., {et~al.} 2025, \aap, 693,
  A288

\bibitem[{{Rodrigo} \& {Solano}(2020)}]{svo2020}
{Rodrigo}, C. \& {Solano}, E. 2020, in XIV.0 Scientific Meeting (virtual) of
  the Spanish Astronomical Society, 182

\bibitem[{{Rodrigo} {et~al.}(2012){Rodrigo}, {Solano}, \& {Bayo}}]{svo2012}
{Rodrigo}, C., {Solano}, E., \& {Bayo}, A. 2012, {SVO Filter Profile Service
  Version 1.0}, IVOA Working Draft 15 October 2012

\bibitem[{{Schlafly} {et~al.}(2019){Schlafly}, {Meisner}, \&
  {Green}}]{Schlafly2019}
{Schlafly}, E.~F., {Meisner}, A.~M., \& {Green}, G.~M. 2019, \apjs, 240, 30

\bibitem[{{Simons} \& {Tokunaga}(2002)}]{Simons2002}
{Simons}, D.~A. \& {Tokunaga}, A. 2002, \pasp, 114, 169

\bibitem[{{Skrutskie} {et~al.}(2006){Skrutskie}, {Cutri}, {Stiening},
  {Weinberg}, {Schneider}, {Carpenter}, {Beichman}, {Capps}, {Chester},
  {Elias}, {Huchra}, {Liebert}, {Lonsdale}, {Monet}, {Price}, {Seitzer},
  {Jarrett}, {Kirkpatrick}, {Gizis}, {Howard}, {Evans}, {Fowler}, {Fullmer},
  {Hurt}, {Light}, {Kopan}, {Marsh}, {McCallon}, {Tam}, {Van Dyk}, \&
  {Wheelock}}]{Skrutskie2006}
{Skrutskie}, M.~F., {Cutri}, R.~M., {Stiening}, R., {et~al.} 2006, \aj, 131,
  1163

\bibitem[{{Smith} {et~al.}(1993){Smith}, {Sellgren}, \& {Brooke}}]{Smith1993}
{Smith}, R.~G., {Sellgren}, K., \& {Brooke}, T.~Y. 1993, \mnras, 263, 749

\bibitem[{{Smith} {et~al.}(2025){Smith}, {Dickinson}, {Fraser}, {McClure},
  {Noble}, {Boogert}, {Sun}, {Egami}, {Dartois}, {Erkal}, {Shimonishi}, {Beck},
  {Bergner}, {Caselli}, {Charnley}, {Chu}, {Drozdovskaya}, {Garrod}, {Harsono},
  {Ioppolo}, {Jimenez-Serra}, {J{\o}rgensen}, {Melnick}, {{\~A}-berg},
  {Palumbo}, {Pendleton}, {Perotti}, {Pontoppidan}, {Qasim}, {Rocha}, {Sturm},
  {Taillard}, {Urso}, \& {van Dishoeck}}]{Smith2025}
{Smith}, Z.~L., {Dickinson}, H.~J., {Fraser}, H.~J., {et~al.} 2025, Nature
  Astronomy

\bibitem[{{Swiggum} {et~al.}(2024){Swiggum}, {Alves}, {Benjamin},
  {Ratzenb{\"o}ck}, {Miret-Roig}, {Gro{\ss}schedl}, {Meingast}, {Goodman},
  {Konietzka}, {Zucker}, {Hunt}, \& {Reffert}}]{Swiggum2024}
{Swiggum}, C., {Alves}, J., {Benjamin}, R., {et~al.} 2024, \nat, 631, 49

\bibitem[{{Taurus Team}(2020)}]{Taurus2020}
{Taurus Team}. 2020, {Taurus 2: Finishing the Spitzer Map of the Taurus
  Molecular Clouds}, NASA IPAC DataSet, IRSA434

\bibitem[{{Tielens} {et~al.}(1984){Tielens}, {Allamandola}, {Bregman},
  {Goebel}, {D'Hendecourt}, \& {Witteborn}}]{Tielens1984}
{Tielens}, A.~G.~G.~M., {Allamandola}, L.~J., {Bregman}, J., {et~al.} 1984,
  \apj, 287, 697

\bibitem[{{Vernet} {et~al.}(2011){Vernet}, {Dekker}, {D'Odorico}, {Kaper},
  {Kjaergaard}, {Hammer}, {Randich}, {Zerbi}, {Groot}, {Hjorth}, {Guinouard},
  {Navarro}, {Adolfse}, {Albers}, {Amans}, {Andersen}, {Andersen}, {Binetruy},
  {Bristow}, {Castillo}, {Chemla}, {Christensen}, {Conconi}, {Conzelmann},
  {Dam}, {de Caprio}, {de Ugarte Postigo}, {Delabre}, {di Marcantonio},
  {Downing}, {Elswijk}, {Finger}, {Fischer}, {Flores}, {Fran{\c{c}}ois},
  {Goldoni}, {Guglielmi}, {Haigron}, {Hanenburg}, {Hendriks}, {Horrobin},
  {Horville}, {Jessen}, {Kerber}, {Kern}, {Kiekebusch}, {Kleszcz}, {Klougart},
  {Kragt}, {Larsen}, {Lizon}, {Lucuix}, {Mainieri}, {Manuputy}, {Martayan},
  {Mason}, {Mazzoleni}, {Michaelsen}, {Modigliani}, {Moehler}, {M{\o}ller},
  {Norup S{\o}rensen}, {N{\o}rregaard}, {P{\'e}roux}, {Patat}, {Pena}, {Pragt},
  {Reinero}, {Rigal}, {Riva}, {Roelfsema}, {Royer}, {Sacco}, {Santin},
  {Schoenmaker}, {Spano}, {Sweers}, {Ter Horst}, {Tintori}, {Tromp}, {van
  Dael}, {van der Vliet}, {Venema}, {Vidali}, {Vinther}, {Vola}, {Winters},
  {Wistisen}, {Wulterkens}, \& {Zacchei}}]{Vernet2011}
{Vernet}, J., {Dekker}, H., {D'Odorico}, S., {et~al.} 2011, \aap, 536, A105

\bibitem[{Virtanen {et~al.}(2020)Virtanen, Gommers, Oliphant, Haberland, Reddy,
  Cournapeau, Burovski, Peterson, Weckesser, Bright, {van der Walt}, Brett,
  Wilson, Millman, Mayorov, Nelson, Jones, Kern, Larson, Carey, Polat, Feng,
  Moore, {VanderPlas}, Laxalde, Perktold, Cimrman, Henriksen, Quintero, Harris,
  Archibald, Ribeiro, Pedregosa, {van Mulbregt}, \& {SciPy 1.0
  Contributors}}]{scipy}
Virtanen, P., Gommers, R., Oliphant, T.~E., {et~al.} 2020, Nature Methods, 17,
  261

\bibitem[{{Weingartner} \& {Draine}(2001)}]{Weingartner2001}
{Weingartner}, J.~C. \& {Draine}, B.~T. 2001, \apj, 548, 296

\bibitem[{{Wenger} {et~al.}(2000){Wenger}, {Ochsenbein}, {Egret}, {Dubois},
  {Bonnarel}, {Borde}, {Genova}, {Jasniewicz}, {Lalo{\"e}}, {Lesteven}, \&
  {Monier}}]{simbad}
{Wenger}, M., {Ochsenbein}, F., {Egret}, D., {et~al.} 2000, \aaps, 143, 9

\bibitem[{{Werner} {et~al.}(2004){Werner}, {Roellig}, {Low}, {Rieke}, {Rieke},
  {Hoffmann}, {Young}, {Houck}, {Brandl}, {Fazio}, {Hora}, {Gehrz}, {Helou},
  {Soifer}, {Stauffer}, {Keene}, {Eisenhardt}, {Gallagher}, {Gautier}, {Irace},
  {Lawrence}, {Simmons}, {Van Cleve}, {Jura}, {Wright}, \&
  {Cruikshank}}]{Werner2004}
{Werner}, M.~W., {Roellig}, T.~L., {Low}, F.~J., {et~al.} 2004, \apjs, 154, 1

\bibitem[{{Wright} {et~al.}(2010){Wright}, {Eisenhardt}, {Mainzer}, {Ressler},
  {Cutri}, {Jarrett}, {Kirkpatrick}, {Padgett}, {McMillan}, {Skrutskie},
  {Stanford}, {Cohen}, {Walker}, {Mather}, {Leisawitz}, {Gautier}, {McLean},
  {Benford}, {Lonsdale}, {Blain}, {Mendez}, {Irace}, {Duval}, {Liu}, {Royer},
  {Heinrichsen}, {Howard}, {Shannon}, {Kendall}, {Walsh}, {Larsen}, {Cardon},
  {Schick}, {Schwalm}, {Abid}, {Fabinsky}, {Naes}, \& {Tsai}}]{Wright2010}
{Wright}, E.~L., {Eisenhardt}, P. R.~M., {Mainzer}, A.~K., {et~al.} 2010, \aj,
  140, 1868

\bibitem[{{Zari} {et~al.}(2016){Zari}, {Lombardi}, {Alves}, {Lada}, \&
  {Bouy}}]{Zari2016}
{Zari}, E., {Lombardi}, M., {Alves}, J., {Lada}, C.~J., \& {Bouy}, H. 2016,
  \aap, 587, A106

\bibitem[{{Zhang} \& {Kainulainen}(2022)}]{Zhang2022}
{Zhang}, M. \& {Kainulainen}, J. 2022, \mnras, 517, 5180

\bibitem[{{Zhang} \& {Green}(2025)}]{Zhang2025}
{Zhang}, X. \& {Green}, G.~M. 2025, Science, 387, 1209

\end{thebibliography}

\begin{appendix}

\section{Supplementary figures and tables}

In this Appendix, I provide supplementary material that supports and extends the analysis presented in the main body of the manuscript. I include additional figures that illustrate key methodological details, data quality assessments, and aspects of the empirical calibration of the \ice method. Furthermore, I also include supplementary tables with reference values for intrinsic stellar colors and the full calibration sample used in this study. 

\begin{figure}
    \centering
    \resizebox{1.0\hsize}{!}{\includegraphics[]{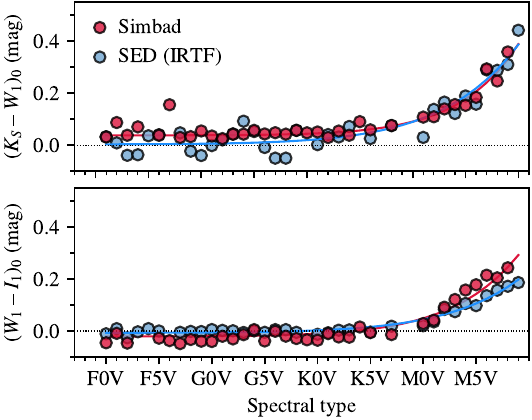}}
    \caption{Comparison of intrinsic color indices derived from synthetic photometry of IRTF spectra (blue points) and from the Simbad database (red points) for dwarf stars. The upper panel shows \(K_S - W_1\) and the lower panel \(W_1 - I_1\) as a function of spectral type. The solid lines indicate power-law fits to the distributions.}
    \label{fig:intrinsic_colors_sed_vs_simbad}
\end{figure}

Figure~\ref{fig:intrinsic_colors_sed_vs_simbad} compares intrinsic color indices derived from synthetic photometry of IRTF spectra with those obtained from the Simbad database for dwarf stars with \textit{Gaia} parallaxes limited to \(\varpi > \SI{1}{mas}\) to mitigate reddening from dust extinction. The figure shows \(K_S - W_1\) colors in the top panel and \(W_1 - I_1\) colors in the bottom panel. Red filled circles represent stellar colors obtained from querying spectral types of dwarf stars as listed in the Simbad database and cross-matching with unWISE, SESNA, and 2MASS. Blue circles denote colors obtained from synthetic photometry on spectra obtained from the IRTF spectral library. Additionally, I computed power-law fits, displayed as red and blue solid lines, to interpolate colors for each spectral type. The excellent agreement between these two approaches validates the use of synthetic colors as intrinsic references for the ice color excess method and supports the robustness of the calibration across a range of spectral types.

\begin{figure}[t]
    \centering
    \resizebox{1.0\hsize}{!}{\includegraphics[]{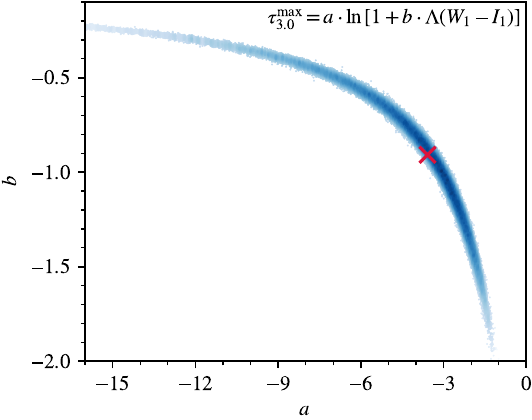}}
    \caption{Distribution of best-fit parameters \(a\) and \(b\) in the empirical calibration of the ice color excess method, specifically \lambdaWISEminusIRAC (see Eq.~\eqref{equ:ice_fit_func}), based on \(10^5\) random fitting realizations. The color scale indicates the density of solutions with darker shades of blue indicating higher density. The geometric median is marked by a red cross and is adopted as the best-fit solution.}
    \label{fig:ice_fit_error}
\end{figure}

Figure~\ref{fig:ice_fit_error} presents the distribution of best-fit parameters obtained from \(10^5\) Monte Carlo realizations of the empirical calibration procedure for fitting \taumax as a function of \lambdaWISEminusIRAC. This was achieved by randomly sampling normal distributions for the input data points, where the shape of each normal distribution was given by the individual mean and error values for \taumax and \lambdaWISEminusIRAC. The results for individual realizations are plotted as small blue dots where darker shades of blue correspond to larger local densities in the displayed parameter space. The geometric median, indicated by a red cross, is adopted as the best-fit solution and demonstrates the stability of the fit against input uncertainties.

\begin{figure*}
    \centering
    \resizebox{1.0\hsize}{!}{\includegraphics[]{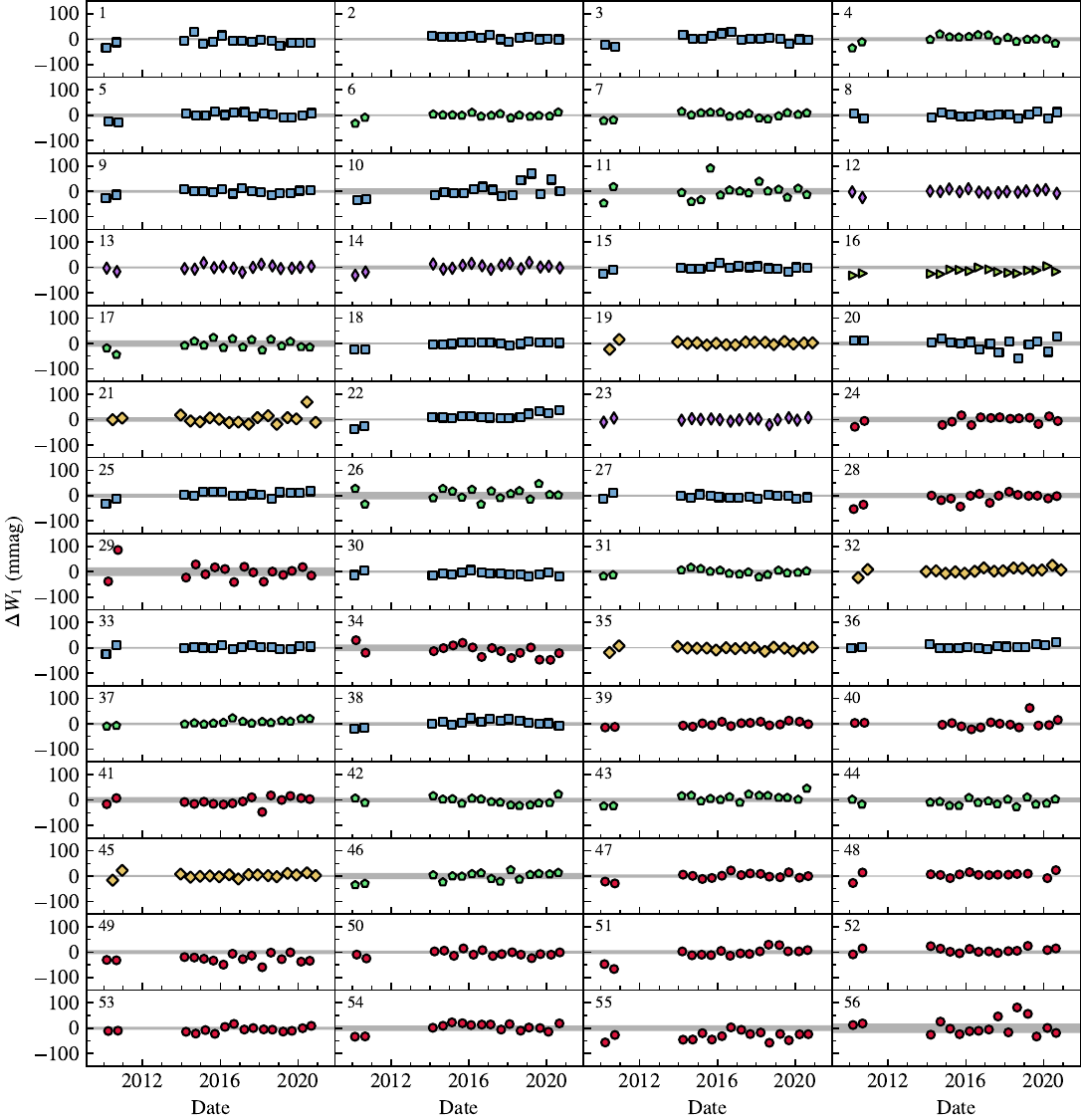}}
    \caption{Time series light curves in the unWISE \(W_1\) band for all 56 sources in the calibration sample. Each panel shows the difference between the unTimely and the unWISE mean magnitude in the \(W_1\) filter as a function of the observing date, demonstrating the photometric variability over the decade-long baseline. The shaded region indicates the inter-quartile range used for the variability quality cut. The source identification number in the top left of each subplot refers to Table~\ref{tab:master} and the symbol colors and shapes are those listed in Table~\ref{tab:literature_overview}.}
    \label{fig:lightcurves_untimely}
\end{figure*}

Figure~\ref{fig:lightcurves_untimely} displays all unWISE \(W_1\) light curves for all 56 sources in the calibration sample used to perform the fit between \taumax and \lambdaWISEminusIRAC. In each panel, the source identification number (as listed in Table~\ref{tab:master}) is indicated in the top left. Data points are represented by their literature-specific colors and symbols as listed in Table.~\ref{tab:literature_overview}. The inter-quartile range (IQR), shown as a shaded region, is used as a quantitative quality criterion to exclude variable sources from the calibration sample.

Table~\ref{tab:intrinsic_colors} lists the intrinsic infrared colors for a range of spectral types, derived from synthetic photometry of IRTF spectra. These values are crucial for accurately determining the color excess attributable to interstellar ices, as discussed in Sect.~\ref{sec:errors_intrinsic_colors}. The table lists intrinsic colors for spectral type F0 through M9 for the colors \(K_S - W_1\), \(W_1 - I_1\), and \(W_1 - L'\). Each color is divided into two sub-columns where the left value corresponds to colors for dwarfs (luminosity class V) and the value on the right corresponds to giants (luminosity class III).

Table~\ref{tab:master} presents the master calibration sample which contains information for the 56 sources used to fit \taumax as a function of \lambdaWISEminusIRAC. For each star, the table provides coordinates, photometric measurements, adopted spectral type, extinction values, measured ice color excess, and literature references. 

\begin{table}
\caption{Intrinsic infrared colors for dwarfs and giants, derived from synthetic photometry of IRTF spectra. Values are given for spectral types F0 to M9 for the color indices \(K_S - W_1\), \(W_1 - I_1\), and \(W_1 - L'\). For each color, the left column lists values for dwarf stars, the right column for giants.}
\centering
\begin{tabular}{@{\extracolsep{\fill}} l c c c}
\hline\hline
& $K_S - W_1$ & $W_1 - I_1$ & $W_1 - L'$ \\
& (mag) & (mag) & (mag)  \\
\hline
F0 & 0.00 / 0.01 & 0.00 / 0.01 & 0.00 / 0.01 \\
F5 & 0.00 / 0.01 & 0.00 / 0.01 & 0.00 / 0.01 \\
G0 & 0.01 / 0.01 & 0.00 / 0.01 & 0.00 / 0.01 \\
G5 & 0.01 / 0.01 & 0.00 / 0.01 & 0.00 / 0.01 \\
K0 & 0.02 / 0.01 & 0.01 / 0.01 & 0.01 / 0.01 \\
K3 & 0.04 / 0.01 & 0.02 / 0.01 & 0.02 / 0.01 \\
K6 & 0.06 / 0.01 & 0.03 / 0.01 & 0.04 / 0.01 \\
K9 & 0.09 / 0.01 & 0.04 / 0.01 & 0.07 / 0.01 \\
M0 & 0.11 / 0.01 & 0.05 / 0.01 & 0.09 / 0.01 \\
M1 & 0.13 / 0.01 & 0.06 / 0.02 & 0.11 / 0.01 \\
M2 & 0.15 / 0.01 & 0.07 / 0.02 & 0.13 / 0.02 \\
M3 & 0.17 / 0.01 & 0.08 / 0.03 & 0.16 / 0.02 \\
M4 & 0.19 / 0.02 & 0.10 / 0.04 & 0.19 / 0.04 \\
M5 & 0.22 / 0.04 & 0.11 / 0.06 & 0.23 / 0.06 \\
M6 & 0.26 / 0.07 & 0.13 / 0.10 & 0.28 / 0.12 \\
M7 & 0.29 / 0.15 & 0.15 / 0.14 & 0.33 / 0.22 \\
M8 & 0.34 / 0.31 & 0.17 / 0.22 & 0.40 / 0.42 \\
M9 & 0.39 / 0.66 & 0.20 / 0.33 & 0.48 / 0.79 \\
\hline
\end{tabular}
\label{tab:intrinsic_colors}
\end{table}

\pagebreak
\pagebreak

\onecolumn
\begin{landscape}
\subsection{Rotated long tables in appendices}

\begin{longtable}{@{\extracolsep{\fill}} l c c c c c c c c c c c c c c c c c}
\caption{Data for the calibration sample of 56 sources that were used for the empirical fit of the ice color excess method. For each source, the table lists coordinates in Right Ascension and Declination (RA, DEC), source magnitudes and errors in the bands \(W_1\) and \(I_1\), the source of the \textit{Spitzer} \(I_1\) measurement (S=SESNA, T=\textit{Spitzer} Taurus, C=c2d), the inter-quartile range for \(W_1\) (\(W_{1,\rm IQR}\)), the adopted extinction and error (\(A_{K_S}\)), the adopted value and error for the peak optical depth (\taumax), the computed value for \lambdaWISEminusIRAC, the adopted spectral type (SpT), and the literature reference for the given measurement with references quoted at the bottom of the table.}\\

\label{tab:master}
ID & RA & Dec & $W_1$ & $\sigma_{W_1}$ & $I_1$ & $I_{1,\rm ref.}$ & $\sigma_{I_1}$ & $W_{1,\rm IQR}$ & $\left( W_1 - I_1 \right)_0$ & $A_{K_S}$ & $\sigma_{A_{K_S}}$ & \taumax & $\sigma_{\taumax}$ & $\Lambda_{W_1I_1}$ & $\sigma_{\Lambda_{W_1I_1}}$ & SpT & Ref. \\
& (hh:mm:ss) & (dd:mm:ss) & (mag) & (mmag) & (mag) & & (mmag) & (mmag) & (mag) & (mag) & (mag) & (mag) & (mag) & (mag) & (mag) & & \\
\hline
\endfirsthead
\caption{continued.}\\
ID & RA & Dec & $W_1$ & $\sigma_{W_1}$ & $I_1$ & $I_{1,\rm ref.}$ & $\sigma_{I_1}$ & $W_{1,\rm IQR}$ & $\left( W_1 - I_1 \right)_0$ & $A_{K_S}$ & $\sigma_{A_{K_S}}$ & \taumax & $\sigma_{\taumax}$ & $\Lambda_{W_1I_1}$ & $\sigma_{\Lambda_{W_1I_1}}$ & SpT & Ref. \\
& (hh:mm:ss) & (dd:mm:ss) & (mag) & (mmag) & (mag) & & (mmag) & (mmag) & (mag) & (mag) & (mag) & (mag) & (mag) & (mag) & (mag) & & \\
\hline
\endhead
\hline
\endfoot
1 & 03:36:08.84 & 31:18:37.6 & 8.1719 & 0.2 & 8.041 & S & 14.0 & 9.0 & 0.064 & 0.00 & 0.05 & 0.09 & 0.05 & 0.067 & 0.026 & M2V & f \\
2 & 03:24:56.06 & 30:26:00.6 & 8.1849 & 0.2 & 8.133 & S & 14.0 & 9.7 & 0.008 & 0.34 & 0.07 & 0.00 & 0.12 & 0.037 & 0.026 & K3III & f \\
3 & 18:29:59.43 & 00:41:00.7 & 8.1951 & 0.2 & 7.987 & S & 14.0 & 9.6 & 0.008 & 1.91 & 0.20 & 0.50 & 0.05 & 0.157 & 0.027 & G4III & f \\
4 & 16:00:00.69 & -42:04:10.1 & 8.2239 & 0.2 & 8.185 & S & 14.0 & 15.3 & 0.008 & 0.41 & 0.03 & 0.11 & 0.02 & 0.022 & 0.029 & K3III & d \\
5 & 18:30:12.20 & 01:15:34.1 & 8.2558 & 0.2 & 8.128 & S & 14.0 & 13.8 & 0.008 & 1.61 & 0.10 & 0.29 & 0.05 & 0.084 & 0.028 & K0III & f \\
6 & 16:00:55.59 & -41:59:59.3 & 8.2628 & 0.2 & 8.247 & S & 14.0 & 6.7 & 0.008 & 0.31 & 0.05 & 0.06 & 0.03 & 0.001 & 0.025 & K7III & d \\
7 & 16:01:44.26 & -41:59:36.6 & 8.2723 & 0.2 & 8.188 & S & 14.0 & 15.7 & 0.008 & 0.67 & 0.03 & 0.16 & 0.03 & 0.062 & 0.029 & K5III & d \\
8 & 18:28:40.39 & 00:44:50.4 & 8.2812 & 0.2 & 8.124 & S & 14.0 & 11.0 & 0.020 & 1.12 & 0.17 & 0.24 & 0.12 & 0.112 & 0.027 & M3III & f \\
9 & 03:29:05.08 & 30:22:07.8 & 8.2993 & 0.2 & 8.169 & S & 14.0 & 11.1 & 0.009 & 0.75 & 0.07 & 0.39 & 0.08 & 0.104 & 0.027 & M0III & f \\
10 & 18:28:52.67 & 00:28:24.1 & 8.3283 & 0.2 & 7.796 & S & 14.0 & 24.3 & 0.080 & 4.75 & 0.44 & 1.63 & 0.06 & 0.345 & 0.036 & M6III & f \\
11 & 16:02:40.89 & -42:03:29.5 & 8.3641 & 0.2 & 8.262 & S & 14.0 & 26.1 & 0.080 & 0.31 & 0.03 & 0.10 & 0.00 & 0.015 & 0.036 & M6III & d \\
12 & 04:42:18.43 & 25:16:53.2 & 8.4249 & 0.3 & 8.377 & T & 7.0 & 6.1 & 0.008 & 0.68 & 0.06 & 0.13 & 0.05 & 0.025 & 0.022 & K1III & a \\
13 & 04:42:53.98 & 25:44:18.0 & 8.4302 & 0.3 & 8.436 & T & 8.0 & 8.1 & 0.008 & 0.24 & 0.09 & 0.03 & 0.05 & -0.019 & 0.023 & G4III & a \\
14 & 04:45:18.02 & 26:16:38.0 & 8.4422 & 0.3 & 8.232 & T & 5.0 & 15.5 & 0.008 & 0.09 & 0.06 & 0.04 & 0.05 & 0.201 & 0.026 & G4III & a \\
15 & 03:45:07.96 & 32:04:01.8 & 8.6398 & 0.3 & 8.595 & S & 14.0 & 7.2 & -0.004 & 0.61 & 0.11 & 0.09 & 0.09 & 0.035 & 0.026 & G3V & f \\
16 & 17:28:05.37 & -26:27:05.3 & 8.6658 & 0.3 & 8.557 & S & 14.0 & 14.0 & 0.020 & 0.96 & 0.10 & 0.25 & 0.02 & 0.067 & 0.028 & M3III & e \\
17 & 16:02:21.29 & -41:58:47.9 & 8.6720 & 0.3 & 8.577 & S & 14.0 & 25.5 & 0.011 & 0.66 & 0.05 & 0.27 & 0.02 & 0.069 & 0.035 & M1III & d \\
18 & 03:45:02.07 & 31:41:19.7 & 8.7089 & 0.3 & 8.705 & S & 14.0 & 7.2 & 0.008 & 0.71 & 0.10 & 0.13 & 0.08 & -0.020 & 0.026 & K2III & f \\
19 & 21:47:22.04 & 47:34:41.0 & 8.7172 & 0.3 & 8.302 & S & 14.0 & 9.1 & 0.008 & 2.15 & 0.21 & 1.45 & 0.14 & 0.359 & 0.026 & K3III & c \\
20 & 18:29:17.00 & 00:37:19.1 & 8.7844 & 0.3 & 8.539 & S & 14.0 & 17.0 & 0.030 & 2.03 & 0.09 & 0.63 & 0.20 & 0.169 & 0.030 & M4III & f \\
21 & 21:45:07.73 & 47:31:15.2 & 8.8017 & 0.3 & 8.655 & S & 14.0 & 17.6 & 0.008 & 0.77 & 0.08 & 0.31 & 0.03 & 0.122 & 0.030 & K6III & c \\
22 & 03:29:36.53 & 31:29:46.5 & 8.8515 & 0.3 & 8.619 & S & 14.0 & 8.4 & 0.008 & 1.63 & 0.05 & 0.55 & 0.05 & 0.188 & 0.026 & K3III & f \\
23 & 04:38:39.29 & 25:51:06.2 & 8.8943 & 0.3 & 8.782 & T & 10.0 & 5.5 & 0.008 & 0.74 & 0.06 & 0.13 & 0.10 & 0.088 & 0.023 & K1III & a \\
24 & 19:21:44.81 & 11:21:20.2 & 8.9614 & 0.3 & 8.644 & C & 61.0 & 21.1 & 0.080 & 2.10 & 0.06 & 0.85 & 0.04 & 0.189 & 0.068 & M6III & b \\
25 & 03:28:10.36 & 30:26:34.1 & 8.9640 & 0.3 & 8.797 & S & 14.0 & 15.2 & 0.008 & 1.27 & 0.09 & 0.49 & 0.06 & 0.131 & 0.029 & G6III & f \\
26 & 16:00:35.35 & -42:09:33.7 & 9.0996 & 0.4 & 8.997 & S & 14.0 & 29.1 & 0.011 & 0.60 & 0.09 & 0.18 & 0.09 & 0.078 & 0.038 & M1III & d \\
27 & 03:26:13.56 & 30:29:22.2 & 9.2117 & 0.4 & 9.019 & S & 14.0 & 8.3 & 0.008 & 0.92 & 0.08 & 0.49 & 0.17 & 0.164 & 0.026 & G9III & f \\
28 & 18:16:52.96 & -18:01:28.9 & 9.2251 & 0.4 & 8.914 & C & 63.0 & 20.5 & 0.011 & 3.00 & 0.18 & 0.96 & 0.10 & 0.232 & 0.069 & M1III & b \\
29 & 18:17:26.90 & -04:38:40.6 & 9.2324 & 0.4 & 8.604 & C & 56.0 & 34.9 & 0.020 & 4.60 & 0.14 & 2.55 & 0.13 & 0.504 & 0.069 & M3III & b \\
30 & 03:42:09.95 & 31:44:13.8 & 9.2442 & 0.4 & 9.164 & S & 14.0 & 9.1 & -0.007 & 0.98 & 0.05 & 0.18 & 0.06 & 0.065 & 0.026 & F0V & f \\
31 & 15:42:36.99 & -34:07:36.3 & 9.2459 & 0.4 & 9.002 & S & 14.0 & 15.3 & 0.008 & 1.79 & 0.09 & 0.83 & 0.02 & 0.196 & 0.029 & G8III & d \\
32 & 21:46:11.63 & 47:34:54.0 & 9.2832 & 0.4 & 8.892 & S & 14.0 & 8.0 & 0.008 & 1.86 & 0.19 & 1.17 & 0.12 & 0.341 & 0.026 & G8III & c \\
33 & 03:33:24.17 & 31:17:47.0 & 9.3772 & 0.4 & 9.275 & S & 14.0 & 6.4 & 0.008 & 1.43 & 0.05 & 0.29 & 0.11 & 0.062 & 0.025 & K1III & f \\
34 & 17:11:15.01 & -27:26:18.2 & 9.6162 & 0.5 & 9.162 & S & 14.0 & 26.1 & 0.080 & 3.91 & 0.12 & 1.39 & 0.07 & 0.286 & 0.036 & M6III & b \\
35 & 21:44:32.93 & 47:34:56.9 & 9.6171 & 0.5 & 9.308 & S & 14.0 & 3.9 & 0.008 & 1.57 & 0.16 & 0.82 & 0.08 & 0.266 & 0.025 & K0III & c \\
36 & 03:30:12.40 & 31:44:40.7 & 9.6296 & 0.5 & 9.482 & S & 14.0 & 9.4 & 0.008 & 1.32 & 0.14 & 0.24 & 0.08 & 0.110 & 0.026 & K0III & f \\
37 & 16:01:28.25 & -41:53:52.2 & 9.6824 & 0.5 & 9.480 & S & 15.0 & 9.0 & 0.008 & 1.57 & 0.12 & 0.53 & 0.03 & 0.159 & 0.027 & K7III & d \\
38 & 03:30:04.75 & 30:23:03.2 & 9.8663 & 0.5 & 9.668 & S & 14.0 & 14.2 & 0.008 & 1.41 & 0.05 & 0.52 & 0.05 & 0.159 & 0.028 & G3III & f \\
39 & 18:16:06.01 & -02:25:54.0 & 9.9055 & 0.5 & 9.726 & C & 57.0 & 12.2 & 0.009 & 1.25 & 0.04 & 0.63 & 0.09 & 0.142 & 0.062 & M0III & b \\
40 & 19:20:15.97 & 11:35:14.6 & 9.9396 & 0.6 & 9.634 & C & 55.0 & 12.1 & 0.011 & 3.14 & 0.09 & 0.79 & 0.04 & 0.224 & 0.060 & M1III & b \\
41 & 15:42:15.46 & -52:48:14.7 & 9.9907 & 0.6 & 9.729 & C & 61.0 & 22.6 & 0.009 & 1.94 & 0.06 & 0.70 & 0.00 & 0.208 & 0.068 & M0III & b \\
42 & 16:01:06.44 & -42:02:02.2 & 10.0278 & 0.6 & 9.801 & S & 14.0 & 17.7 & 0.008 & 1.91 & 0.07 & 0.74 & 0.04 & 0.176 & 0.030 & K0III & d \\
43 & 16:01:42.52 & -41:53:06.5 & 10.0567 & 0.6 & 9.676 & S & 14.0 & 17.3 & 0.011 & 2.46 & 0.10 & 1.03 & 0.03 & 0.314 & 0.030 & M1III & d \\
44 & 16:01:26.36 & -41:50:42.3 & 10.0620 & 0.6 & 9.757 & S & 14.0 & 19.2 & 0.009 & 1.65 & 0.08 & 0.71 & 0.04 & 0.258 & 0.031 & M0III & d \\
45 & 21:47:35.09 & 47:37:16.4 & 10.0863 & 0.6 & 10.007 & S & 14.0 & 8.0 & 0.008 & 0.67 & 0.07 & 0.24 & 0.02 & 0.057 & 0.026 & K1III & c \\
46 & 16:00:47.40 & -42:03:57.4 & 10.1197 & 0.6 & 9.854 & S & 14.0 & 24.0 & 0.008 & 2.03 & 0.08 & 0.63 & 0.03 & 0.212 & 0.034 & K5III & d \\
47 & 18:14:07.12 & -07:08:41.4 & 10.1424 & 0.6 & 9.876 & C & 55.0 & 13.2 & 0.008 & 1.89 & 0.06 & 0.83 & 0.04 & 0.216 & 0.060 & K7III & b \\
48 & 17:15:55.74 & -20:55:31.3 & 10.2758 & 0.7 & 9.958 & C & 54.0 & 4.6 & 0.011 & 2.65 & 0.08 & 0.97 & 0.05 & 0.246 & 0.058 & M1III & b \\
49 & 15:42:16.98 & -52:47:43.9 & 10.5690 & 0.8 & 10.183 & C & 55.0 & 15.6 & 0.008 & 2.87 & 0.09 & 0.85 & 0.25 & 0.313 & 0.061 & K7III & b \\
50 & 18:17:04.70 & -08:14:49.5 & 10.7264 & 0.8 & 10.067 & C & 61.0 & 11.5 & 0.009 & 4.27 & 0.13 & 2.33 & 0.12 & 0.554 & 0.065 & M0III & b \\
51 & 17:11:15.38 & -27:27:14.6 & 10.7844 & 0.9 & 10.339 & S & 14.0 & 15.7 & 0.020 & 3.53 & 0.11 & 1.20 & 0.06 & 0.345 & 0.029 & M3III & b \\
52 & 17:16:04.67 & -20:57:07.3 & 10.8233 & 0.9 & 10.400 & C & 54.0 & 12.9 & 0.008 & 3.24 & 0.10 & 1.27 & 0.06 & 0.342 & 0.059 & K7III & b \\
53 & 18:17:13.67 & -08:13:18.7 & 10.9758 & 0.9 & 10.474 & C & 58.0 & 11.7 & 0.008 & 3.58 & 0.11 & 2.02 & 0.10 & 0.413 & 0.063 & K7III & b \\
54 & 04:21:54.04 & 15:30:29.8 & 10.9794 & 0.9 & 10.575 & C & 53.0 & 21.7 & 0.008 & 3.04 & 0.09 & 1.09 & 0.05 & 0.328 & 0.061 & K0III & b \\
55 & 18:17:11.81 & -08:14:01.4 & 11.3842 & 1.2 & 10.752 & C & 55.0 & 23.1 & 0.008 & 4.31 & 0.13 & 2.26 & 0.11 & 0.527 & 0.063 & K7III & b \\
56 & 17:11:20.05 & -27:27:13.3 & 11.4173 & 1.3 & 10.819 & S & 15.0 & 37.5 & 0.020 & 6.00 & 0.18 & 2.25 & 0.11 & 0.442 & 0.045 & M3III & b \\
\end{longtable}
\tablebib{(a)~\citet{Murakawa2000}; (b) \citet{Boogert2011}; (c) \citet{Chiar2011}; (d) \citet{Boogert2013}; (e) \citet{Goto2018}; (f) \citet{Madden2022}.}

\end{landscape}

\FloatBarrier 
\clearpage

\end{appendix}

\end{document}